\documentclass{emulateapj}
\usepackage{apjfonts}
\journalinfo{The Astrophysical Journal, 689, 2008 December 10, in press}

\shorttitle{ACCRETION-DRIVEN WINDS FROM T TAURI STARS}
\shortauthors{S. R. CRANMER}

\begin{document}

\title{Turbulence-driven Polar Winds from T Tauri Stars
Energized by Magnetospheric Accretion}

\author{Steven R. Cranmer}
\affil{Harvard-Smithsonian Center for Astrophysics,
60 Garden Street, Cambridge, MA 02138 \\
Submitted 2008 July 8; \, accepted 2008 August 15}

\begin{abstract}
Pre-main-sequence stars are observed to be surrounded by both
accretion flows and some kind of wind or jet-like outflow.
Recent work by Matt and Pudritz has suggested that if classical
T Tauri stars exhibit stellar winds with mass loss rates about 0.1
times their accretion rates, the wind can carry away enough angular
momentum to keep the stars from being spun up unrealistically by
accretion.
This paper presents a preliminary set of theoretical models of
accretion-driven winds from the polar regions of T Tauri stars.
These models are based on recently published self-consistent
simulations of the Sun's coronal heating and wind acceleration.
In addition to the convection-driven MHD turbulence (which dominates
in the solar case), we add another source of wave energy at the
photosphere that is driven by the impact of plasma in neighboring
flux tubes undergoing magnetospheric accretion.
This added energy, determined quantitatively from the far-field
theory of MHD wave generation, is sufficient to produce T Tauri-like
mass loss rates of at least 0.01 times the accretion rate.
While still about an order of magnitude below the level required for
efficient angular momentum removal, these are the first self-consistent
models of T Tauri winds that agree reasonably well with a range of
observational mass loss constraints.
The youngest modeled stellar winds are supported by Alfv\'{e}n wave
pressure, they have low temperatures (``extended chromospheres''),
and they are likely to be unstable to the formation of
counterpropagating shocks and clumps far from the star.
\end{abstract}

\keywords{accretion, accretion disks --- stars: coronae ---
stars: mass loss --- stars: pre-main sequence ---
stars: winds, outflows --- turbulence}

\section{Introduction}

Our current state of knowledge about how stars and planets are
formed comes from an intertwined web of observations (spanning the
electromagnetic spectrum) and theoretical work.
The early stages of low-mass star formation comprise a wide
array of inferred physical processes, including disk accretion,
various kinds of outflow, and magnetohydrodynamic (MHD)
activity on time scales ranging from hours to millennia
(see, e.g., Lada 1985; Bertout 1989; Appenzeller \& Mundt 1989;
Hartmann 2000; K\"{o}nigl \& Pudritz 2000; McKee \& Ostriker 2007;
Shu et al.\  2007).
A key recurring issue is that there is a great deal of {\em mutual
interaction and feedback} between the star and its circumstellar
environment.
This interaction can determine how rapidly the star rotates,
how active the star appears from radio to X-ray wavelengths, and
how much mass and energy the star releases into its
interplanetary medium.

A key example of circumstellar feedback is the magnetospheric
accretion paradigm for classical T Tauri stars
(Lynden-Bell \& Pringle 1974; Uchida \& Shibata 1984;
Camenzind 1990; K\"{o}nigl 1991).
Because of strong ($\sim$1 kG) stellar magnetic fields, the evolving
equatorial accretion disk does not penetrate to the stellar surface,
but instead is stopped by the stellar magnetosphere.
Accretion is thought to proceed via ballistic infall along magnetic
flux tubes threading the inner disk, leading to shocks and hot spots
on the surface.
The primordial accretion disk is dissipated gradually as the star
enters the weak-lined T Tauri star phase, with a likely transition
to a protoplanetary dust/debris disk.

Throughout these stages, solar-mass stars are inferred to exhibit
some kind of wind or jet-like outflow.
There are several possible explanations of how and where the outflows
arise, including extended disk winds, X-winds, impulsive
(plasmoid-like) ejections, and ``true'' stellar winds (e.g.,
Paatz \& Camenzind 1996; Calvet 1997; K\"{o}nigl \& Pudritz 2000;
Dupree et al.\  2005; Edwards et al.\  2006; Ferreira et al.\  2006;
G\'{o}mez de Castro \& Verdugo 2007; Cai et al.\  2008).
Whatever their origin, the outflows produce observational
diagnostics that indicate mass loss rates exceeding the Sun's present
mass loss rate by factors of $10^3$ to $10^6$.
It is of some interest to evaluate how much of the observed outflow
can be explained solely with {\em stellar} winds, since these flows
are locked to the star and thus are capable of removing angular
momentum from the system.
Recent work by Matt \& Pudritz (2005, 2007, 2008)
has suggested that if there is a stellar wind with a sustained mass
loss rate about 10\% of the accretion rate, the wind can carry away
enough angular momentum to keep T Tauri stars from being spun up
unrealistically by the accretion.

Despite many years of study, the dominant physical processes that
accelerate winds from cool stars have not yet been identified
conclusively.  
For many stars, the large-scale energetics of the system---i.e., the
luminosity and the gravitational potential---seem to determine the
overall magnitude of the mass loss (Reimers 1975; 1977;
Schr\"{o}der \& Cuntz 2005).
Indeed, for the most luminous cool stars, radiation pressure seems
to provide a direct causal link between the luminosity $L_{\ast}$
and the mass loss rate $\dot{M}_{\rm wind}$ (e.g.,
Gail \& Sedlmayr 1987; H\"{o}fner 2005).
However, for young solar-mass stars, other mediating processes
(such as coronal heating, waves, or time-variable magnetic ejections)
are more likely to connect the properties of the stellar interior
to the surrounding outflowing wind.
For example, magnetohydrodynamic (MHD) waves have been studied for
several decades as a likely way for energy to be transferred
from late-type stars to their winds (Hartmann \& MacGregor 1980;
DeCampli 1981; Airapetian et al.\  2000;
Falceta-Gon\c{c}alves et al.\  2006; Suzuki 2007).

In parallel with the above work in improving our understanding of
stellar winds, there has been a great deal of progress toward
identifying and characterizing the processes that produce the
{\em Sun's} corona and wind.
It seems increasingly clear that closed magnetic loops in the
low solar corona are heated by small-scale, intermittent magnetic
reconnection that is driven by the continual stressing of their
footpoints by convective motions (e.g., Aschwanden 2006; Klimchuk 2006).
The open field lines that connect the Sun to interplanetary space,
though, appear to be energized by the dissipation of waves and
turbulent motions that originate at the stellar surface
(Tu \& Marsch 1995; Cranmer 2002; Suzuki 2006; Kohl et al.\  2006).
Parker's (1958) classic paradigm of solar wind acceleration via
gas pressure in a hot ($T \sim 10^{6}$ K) corona still
seems to be the dominant mechanism, though waves and turbulence
have an impact as well.
A recent self-consistent model of turbulence-driven coronal heating
and solar wind acceleration has succeeded in reproducing a wide
range of observations {\em with no ad-hoc free parameters}
(Cranmer et al.\  2007).
This progress on the solar front is a fruitful jumping-off point
for a better understanding of the basic physics of winds and
accretion in young stars.

The remainder of this paper is organized as follows.
{\S}~2 presents an overview of the general scenario of
accretion-driven MHD waves that is proposed here to be important
for driving T Tauri mass loss.
In {\S}~3 the detailed properties of an evolving solar-mass star
are presented, including the fundamental stellar parameters,
the accretion rate and disk geometry, and the properties of the
clumped gas in the magnetospheric accretion streams.
These clumped streams impact the stellar surface and create
MHD waves that propagate horizontally to the launching points of
stellar wind streams.
{\S}~4 describes how self-consistent models of these wind regions
are implemented, and {\S}~5 gives the results.
Finally, {\S}~6 contains a summary of the major results of
this paper and a discussion of the implications these results
may have on our wider understanding of low-mass star formation.

\section{Basal Wave Driving from Inhomogeneous Accretion}

\begin{figure}
\epsscale{1.11}
\plotone{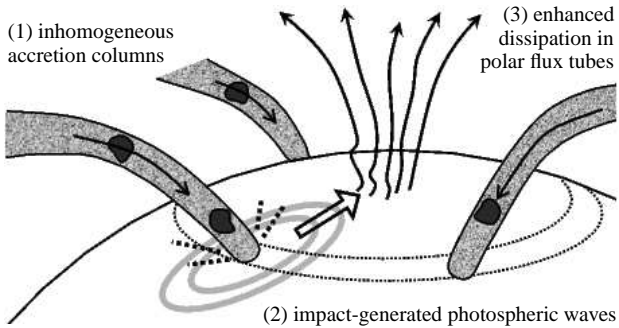}
\caption{Summary sketch of the accretion and wind geometry
discussed in this paper.
(1) Magnetospheric accretion streams are assumed to impact the
star at mid-latitudes.
(2) Dense clumps in the accretion streams generate MHD waves
that propagate horizontally over the stellar surface.
(3) Polar magnetic field lines exhibit enhanced photospheric
wave activity and thus experience stronger coronal heating and
stellar wind acceleration.}
\end{figure}

Figure 1 illustrates the proposed connection between accretion and
stellar wind driving that is explored below.
The models make use of the idea that magnetospheric accretion
streams are likely to be highly unstable and time-variable,
and thus much of the material deposited onto the star is expected to
be in the form of intermittent dense clumps.
The impact of each clump is likely to generate MHD waves on
the stellar surface (Scheurwater \& Kuijpers 1988) and these waves
can propagate across the surface, spread out the kinetic energy
of accretion, and inject some of it into the global magnetic
field.\footnote{%
A somewhat similar wave-generation scenario was suggested by
Vasconcelos et al.\  (2002) and Elfimov et al.\  (2004), but their
model was applied only to the energization of the accretion streams
themselves, not to the regions exterior to the streams.}
The models simulate the time-steady properties of the polar magnetic
flux tubes, wherein the source of wave/turbulence energy at the
photosphere comes from two sources:
(1) accretion-driven waves that have propagated over the surface
from the ``hot spots,'' and
(2) the ever-present sub-photospheric convection, which
shakes the magnetic field lines even without accretion, and is
the dominant source of waves for the present-day Sun.

There is substantial observational evidence that the accretion
streams of classical T Tauri stars are clumped and inhomogeneous
(see, e.g., Gullbring et al.\  1996; Safier 1998;
Bouvier et al.\  2003, 2004, 2007; Stempels 2003).
Strong variations in the accretion signatures take place over
time scales ranging between a few hours (i.e., the free-fall time
from the inner edge of the disk to the star) and a few days (i.e.,
the stellar rotation period).
Some of this variability may arise from instabilities in the
accretion shock (Chevalier \& Imamura 1982; Mignone 2005),
with the possibility of transitions between stable and unstable
periods of accretion (e.g., Romanova et al.\  2008).
Observations of the bases of accretion streams indicate
the presence of turbulence (Johns \& Basri 1995) as well as discrete
events that could signal the presence of magnetic reconnection
(Giardino et al.\  2007).
Alternately, some of the observed variability may come from
larger-scale instabilities in the torqued magnetic field
that is threaded by the disk (e.g., Long et al.\  2007;
Kulkarni \& Romanova 2008) and which could also drive periodic
reconnection events (van Ballegooijen 1994).
The latter variations would not necessarily be limited to the
rotation time scale, since ``instantaneous'' topology changes can
happen much more rapidly.

The mechanisms by which impulsive accretion events (i.e.,
impacts of dense clumps) can create MHD waves on the stellar
surface are described in more detail in {\S}~3.4.
This paper uses the analytic results of
Scheurwater \& Kuijpers (1988) to set the energy flux in these
waves, based on the available kinetic energy in the impacts.
It is useful to point out, however, that there are relatively
well-observed examples of impulse-generated waves on the Sun
that help to justify the overall plausibility of this scenario.
These are the so-called ``Moreton waves'' and ``EIT waves'' that
are driven by solar flares and coronal mass ejections (CMEs).
Moreton waves are localized ripples seen in the H$\alpha$
chromosphere expanding out from strong flares and erupting
filaments at speeds of 500 to 2000 km s$^{-1}$ (Moreton \& Ramsey 1960).
These seem to be consistent with fast-mode MHD shocks associated with
the flaring regions (e.g., Uchida 1968; Balasubramaniam et al.\  2007).
A separate phenomenon, discovered more recently in the extreme
ultraviolet and called EIT waves (Thompson et al.\  1998),
is still not yet well understood.
These waves propagate more slowly than Moreton waves (typically
100 to 400 km s$^{-1}$), but they are often seen to traverse the
entire diameter of the Sun and remain coherent.
Explanations include fast-mode MHD waves (Wang 2000),
solitons (Wills-Davey et al.\  2007), and sheared current layers
containing enhanced Joule heating (Delann\'{e}e et al.\  2008).
These serve as ample evidence that impulsive phenomena can
generate MHD fluctuations that travel across a stellar surface.

Finally, it is important to clarify some of the limitations of the
modeling approach used in this paper.
The models include only a description of the magnetospheric accretion
streams (with an assumed dipole geometry) and the open flux tubes
at the north and south poles of the star.
Thus, there is no attempt to model either disk winds or the
closed-field parts of the stellar corona, despite the fact that
these are likely to contribute to many key observational diagnostics
of T Tauri circumstellar gas.
In addition, the accretion streams themselves are included only for
their net dynamical impact on the polar (non-accreting) regions, and
thus there is no need to describe, e.g., the temperature or ionization
state inside the streams.
The wind acceleration in the polar flux tubes is treated with a
one-fluid, time-steady approximation similar to that described
by Cranmer et al.\  (2007) for the solar wind.
To summarize, this paper is a ``proof of concept'' study to evaluate
how much about T Tauri outflows can be explained with {\em only}
polar stellar winds that are energized by accretion-driven waves.

\section{Baseline Solar-Mass Evolution Model}

\subsection{Stellar Parameters and Accretion Rate}

To begin exploring how accretion-driven waves affect the winds of
young stars, a representative evolutionary sequence of fundamental
parameters was constructed for a solar-type star.
The adopted parameters are not intended to reproduce any specific
observation in detail and should be considered illustrative.
The STARS evolution code\footnote{%
The December 2003 version was obtained from:$\,$
http://www.ast.cam.ac.uk/$\sim$stars/}
(Eggleton 1971, 1972, 1973; Eggleton et al.\  1973;
Han et al.\  1994; Pols et al.\  1995) was used to evolve a
star having a constant mass $M_{\ast} = 1 \, M_{\odot}$ from 
the Hayashi track to well past the current age of the Sun.
Neither mass accretion nor mass loss were included in the
evolutionary calculation, and a standard solar abundance mixture
($X = 0.70$, $Z = 0.02$) was assumed.
All other adjustable parameters of the code were held fixed at
their default values.

\begin{figure}
\epsscale{0.99}
\plotone{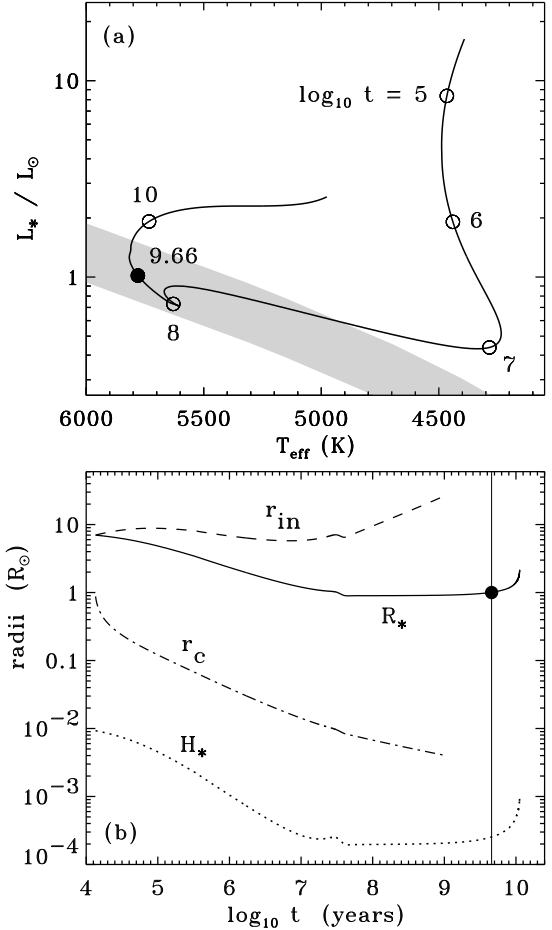}
\caption{Basic properties of 1 $M_{\odot}$ star.
({\em{a}}) H-R diagram showing the main sequence ({\em gray band})
and the modeled evolutionary track, with several representative
ages ({\em open circles}) including the present-day Sun
({\em filled circle}).
({\em{b}}) Length scales, plotted in units of solar radii, as a
function of age: stellar radius ({\em solid line}),
inner edge of accretion disk ({\em dashed line}), accretion
clump radius ({\em dot-dashed line}), and photospheric pressure
scale height ({\em dotted line}).}
\end{figure}

Figure 2 presents a summary of the modeled stellar parameters.
A Hertzsprung-Russell (H-R) diagram is shown in Figure 2{\em{a}},
with the bolometric luminosity $L_{\ast}$ plotted as a function of
the effective temperature $T_{\rm eff}$ for both the evolutionary
model (with a subset of stellar ages $t$ indicated by symbols) and
an approximate main sequence band for luminosity class V stars
(de Jager \& Nieuwenhuijzen 1987).
The current age of the Sun is denoted by $t = 4.6 \times 10^{9}$ yr,
or $\log_{10} t = 9.66$.
Figure 2{\em{b}} shows the age dependence of a selection of
relevant length scales, including the stellar radius $R_{\ast}$
and the photospheric pressure scale height $H_{\ast}$.
The latter quantity is defined assuming a pure hydrogen gas, as
\begin{equation}
  H_{\ast} \, = \, \frac{k_{\rm B} T_{\rm eff}}{m_{\rm H}}
  \, \frac{R_{\ast}^2}{G M_{\ast}}
  \label{eq:Hast}
\end{equation}
where $k_{\rm B}$ is Boltzmann's constant, $m_{\rm H}$ is the
mass of a hydrogen atom, and $G$ is the Newtonian gravitation
constant.
Note that for most of the applications below, only the relative
variation of $H_{\ast}$ with age is needed and not its absolute value.
The scale height is assumed to be proportional to the horizontal
granulation scale length (e.g., Robinson et al.\  2004), which in
turn governs the perpendicular correlation length of Alfv\'{e}nic
turbulence in the photosphere (see {\S}~4).

The other major parameter to be specified as a function of age is
the mass accretion rate $\dot{M}_{\rm acc}$.
For classical T Tauri stars, it is generally accepted that the
accretion rate decreases with increasing stellar age $t$, but there
is a large spread in measured values that may be affected by
observational uncertainties in both $\dot{M}_{\rm acc}$ and $t$.
In order to determine a representative age dependence for
$\dot{M}_{\rm acc}(t)$, we utilized tabulated accretion rates
from Hartigan et al.\  (1995) and Hartmann et al.\  (1998),
who interpreted optical/UV continuum excesses as an ``accretion
luminosity'' that comes from the kinetic energy of accreted gas
impacting the star.
There is some overlap in the star lists from these two sources,
and differences in the diagnostic techniques resulted in different
values of both $\dot{M}_{\rm acc}$ and $t$ for some stars common to
both lists.
Both values have been retained here, and thus Figure 3{\em{a}}
shows a total of 88 data points from both lists.\footnote{%
This figure shows relatively nearby Galactic stars only.
In the LMC, the mass accretion rates seem to be much higher
at ages around 10 Myr (e.g., Romaniello et al.\  2004), which could
indicate a substantial metallicity dependence for many properties of
the T Tauri phase.}
The observational uncertainties are not shown; they may be as large
as an order of magnitude in $\dot{M}_{\rm acc}$ and a factor of
$\sim$3 in the age (see, e.g., Figure 3 of Muzerolle et al.\  2000).

\begin{figure}
\epsscale{1.05}
\plotone{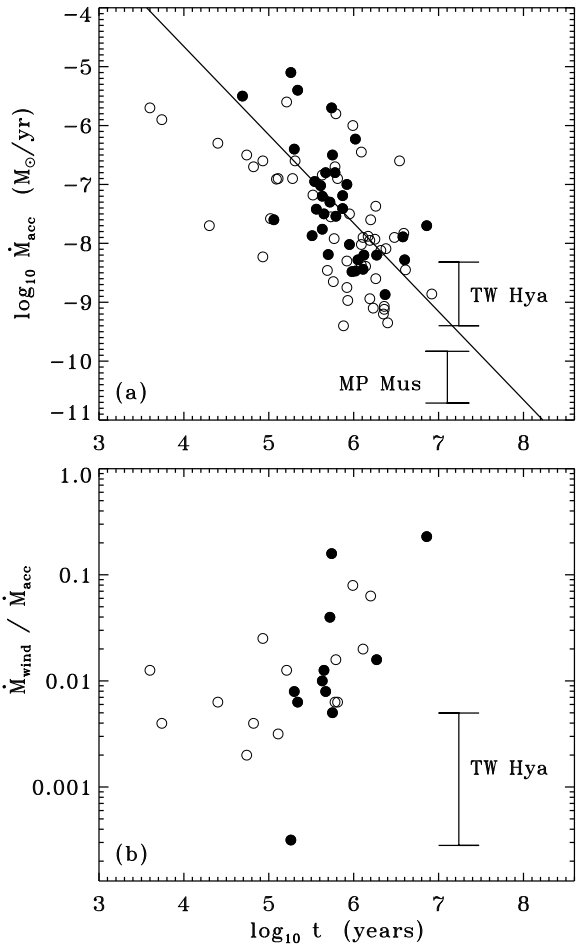}
\caption{Mass accretion rates and mass loss rates for classical
T Tauri stars.
({\em{a}}) Measured accretion rates from Hartigan et al.\  (1995)
and Hartmann et al.\  (1998) ({\em open circles}) with the subset
of roughly solar-mass stars highlighted ({\em filled circles}).
The best-fit curve from eq.~(\ref{eq:Maccfit}) is also shown
({\em solid line}) as well as two measured rates for the stars
TW Hya and MP Mus ({\em error bars}); see text for details.
({\em{b}}) Ratio of mass loss rate to mass accretion rate for
measured T Tauri stars, where the symbols have the same meanings
as in panel ({\em{a}}).}
\end{figure}

Highlighted in Figure 3{\em{a}}, as filled symbols, are a subset
of stars that appear to be on the 1 $M_{\odot}$ Hayashi track (i.e.,
they have $T_{\rm eff}$ between about 4000 and 4500 K).
An initial attempt to fit these 35 stars to a power-law age
dependence yielded
\begin{equation}
  \dot{M}_{\rm acc} \, \approx \, 2.8 \times 10^{-8} \,
  M_{\odot} \, \mbox{yr}^{-1} \, \left( \frac{t}{10^{6} \, \mbox{yr}}
  \right)^{-1.22}  \,\, .
\end{equation}
However, the theoretical accretion disk models of
Hartmann et al.\  (1998) found that the exponent $\eta$ in
$\dot{M}_{\rm acc} \propto t^{-\eta}$ is most likely to range
between about 1.5 and 2.8, with the lower end of that range being
most likely.
Thus, the fitted value of $\eta = 1.22$ above was judged to be
too low, and a more realistic fit was performed by fixing 
$\eta = 1.5$ and finding
\begin{equation}
  \dot{M}_{\rm acc} \, \approx \, 2.2 \times 10^{-8} \,
  M_{\odot} \, \mbox{yr}^{-1} \, \left( \frac{t}{10^{6} \, \mbox{yr}}
  \right)^{-1.5}   \label{eq:Maccfit}
\end{equation}
which is used in the wind models below and is plotted in
Figure 3{\em{a}}.
The scatter around the mean fit curve has a roughly normal
distribution (in the logarithm of $\dot{M}_{\rm acc}$) with a
standard deviation ranging between a factor of 6 (for the 35
solar-mass stars) to 8 (for all 88 data points) on either side of
the representative trend given by equation (\ref{eq:Maccfit}).
The sensitivity of the resulting accretion-driven wind to this
scatter is explored further below in {\S}~5.4.

Classical T Tauri magnetospheric accretion is believed to end at an
age around 10 Myr (e.g., Strom et al.\  1989; Bouvier et al.\  1997).
This is probably coincident with the time when the inner edge of the
accretion disk grows to a larger extent than the Keplerian corotation
radius, with ``propeller-like'' outflow replacing the accretion
(Illarionov \& Sunyaev 1975; Ustyugova et al.\  2006).
Rotation is not included in the models presented here, so this
criterion is not applied directly.
However, for advanced ages (e.g., $t \gtrsim 100$ Myr) equation
(\ref{eq:Maccfit}) gives increasingly weak accretion rates that 
end up not having any significant effect on the stellar wind.
Thus, even though the Figures below plot the various accretion-driven
quantities up to an age of 1 Gyr ($\log_{10} t = 9$), the models
below do not apply any abrupt cutoff to the accretion rate.

Figure 3{\em{b}} plots mass loss rates $\dot{M}_{\rm wind}$ for a
subset of classical T Tauri stars that have measurements of
blueshifted emission in forbidden lines such as [\ion{O}{1}]
$\lambda$6300 (Hartigan et al.\  1995).
The mass loss rates are shown as an efficiency ratio
$\dot{M}_{\rm wind}/\dot{M}_{\rm acc}$ which depends on the combined
uncertainties of both the outflow and accretion properties.
The largest ratio shown, at $\log_{10} t \approx 6.9$, is for
the jet of HN Tau, and the original value from
Hartigan et al.\  (1995) has been replaced with the slightly
lower revised value from Hartigan et al.\  (2004).
It is uncertain whether the measured outflows originate on the
stellar surface or in the accretion disk (see, e.g.,
Paatz \& Camenzind 1996; Calvet 1997; Ferreira et al.\  2006),
but these rates can be used as upper limits for any
stellar wind component.

Figure 3 also gives additional information about two of the oldest
reported classical T Tauri stars: TW Hya and MP Mus.
Note, however, that their extremely uncertain accretion rates
were not included in the fitting for $\dot{M}_{\rm acc} (t)$.
The age of TW Hya is quite uncertain; it is often given as
$\sim$10 Myr (e.g., Muzerolle et al.\  2000), but values as large
as 30 Myr have been computed (Batalha et al.\  2002)
and the plotted error bar attempts to span this range.
Similarly, the age of MP Mus seems to be between 7 and 23 Myr
(e.g., Argiroffi et al.\  2007).
The plotted upper and lower limits on $\dot{M}_{\rm acc}$ for
TW Hya come from from Batalha et al.\  (2002) and
Muzerolle et al.\  (2000), respectively, and the range of values
for $\dot{M}_{\rm wind}$ in Figure 3{\em{b}} were taken from
Dupree et al.\  (2005).
The plotted ratio, however, divides the observational limits on
$\dot{M}_{\rm wind}$ by a mean value of $\dot{M}_{\rm acc} =
1.4 \times 10^{-9}$ $M_{\odot}$ yr$^{-1}$, in order to avoid
creating an unrealistically huge range.
For MP Mus, the mean accretion rate of $5 \times 10^{-11}$
$M_{\odot}$ yr$^{-1}$ was derived from X-ray measurements by
Argiroffi et al.\  (2007).
The plotted error bars for $\dot{M}_{\rm acc}$ were estimated
by using the 1$\sigma$ uncertainties on the X-ray emission measure
and column density.

\subsection{Magnetospheric Accretion Streams}

The dynamical properties of accretion streams are modeled here
using an axisymmetric dipole magnetic field, coupled with the
assumption of ballistic infall (e.g., Calvet \& Hartmann 1992;
Muzerolle et al.\  2001).
Although it is almost certain that the actual magnetic fields of
T Tauri stars are not dipolar (Donati et al.\  2007;
Jardine et al.\  2008), this assumption
allows the properties of the accretion streams to be calculated
simply and straightforwardly as a function of evolutionary age.

Stellar field lines that thread the accretion disk are bounded
between inner and outer radii $r_{\rm in}$ and $r_{\rm out}$ as
measured in the equatorial plane.
We assume the inner radius---also called the ``truncation radius''---is
described by K\"{o}nigl's (1991) application of neutron-star accretion
theory (see also Davidson \& Ostriker 1973; Elsner \& Lamb 1977;
Ghosh \& Lamb 1979a,b).
This expression for $r_{\rm in}$ is given by determining where the
magnetic pressure of the inner dipole region balances the gas pressure
in the outer accretion region.
Assuming free-fall in the accretion stream,
\begin{equation}
  r_{\rm in} \, = \, \beta_{\rm GL} \left(
  \frac{B_{\ast}^{4} R_{\ast}^{12}}
  {2G M_{\ast} \dot{M}_{\rm acc}^{2}} \right)^{1/7}
  \label{eq:K91}
\end{equation}
where the scaling constant $\beta_{\rm GL}$ describes the departure
from ideal magnetostatic balance; i.e., $\beta_{\rm GL} = 1$ gives
the standard ``Alfv\'{e}n radius'' at which the pressures balance.
Following K\"{o}nigl (1991), the value $\beta_{\rm GL} = 0.5$ is
used here.
The outer disk radius $r_{\rm out}$ may be as large as the Keplerian
corotation radius, but it may not extend that far in reality (e.g.,
Hartmann et al.\  1994; Bessolaz et al.\  2008).
For simplicity, the outer disk radius is assumed to scale with the
inner disk radius as
\begin{equation}
  r_{\rm out} \, = \, r_{\rm in} (1 + \epsilon)
\end{equation}
where a constant value of $\epsilon = 0.1$ was adopted.
This value falls within the range of empirically constrained
outer/inner disk ratios used by Muzerolle et al.\  (2001) to model
H$\alpha$ profiles; they used effective values of $\epsilon$ between
0.034 and 0.36.
The intermediate value of 0.1 produces reasonable magnitudes for
the filling factors of accretion stream footpoints on the stellar
surface (see below).

Note that equation (\ref{eq:K91}) above requires the specification
of the surface magnetic field strength $B_{\ast}$.
In the models of the open flux tubes (in the {\em polar} regions)
used below, a solar-type magnetic field is adopted.
The photospheric value of $B_{\ast} \approx 1500$ G
is held fixed for the footpoints of the stellar wind flux tubes
(see Cranmer \& van Ballegooijen 2005).
For the {\em mid-latitude} field at the footpoints of the accretion
streams, though, a slightly weaker field of 1000 G is used.

Figure 2{\em{b}} shows the resulting age dependence for $r_{\rm in}$.
For the youngest modeled ``protostars'' ($t \lesssim 10^{4}$ yr),
the accretion rate is so large that $r_{\rm in}$ would be smaller
than the stellar radius itself.
In that case, the accretion disk would penetrate down to the
star and there would be no magnetospheric infall.
Thus, the youngest age considered for the remainder of this paper
is $t = 13.5$ kyr (i.e., $\log_{10} t = 4.13$), for which
$r_{\rm in} = R_{\ast}$.
Over most of the classical T Tauri age range (0.1--10 Myr),
$r_{\rm in}$ decreases slightly in absolute extent, but the
ratio $r_{\rm in}/R_{\ast}$ increases and is approximately
proportional to $t^{0.2}$.

For any radius $r > R_{\ast}$ in the equatorial disk, an aligned
dipole field line impacts the stellar surface at colatitude $\theta$,
where $R_{\ast} = r \sin^{2}\theta$.
This allows the colatitudes $\theta_{\rm in}$ and $\theta_{\rm out}$
to be computed from $r_{\rm in}$ and $r_{\rm out}$, and these can
be used to compute the fraction of stellar surface area $\delta$
that is filled by accretion streams, with
\begin{equation}
  \delta \, = \, \cos\theta_{\rm out} - \cos\theta_{\rm in}
  \label{eq:delta}
\end{equation}
(see also Lamzin 1999).
Note that both the northern and southern polar ``rings'' are taken
into account in the above expression.
The filling factor $\delta$ depends sensitively on the geometry of
the accretion volume; i.e., for a dipole field, it depends only on
the outer/inner radius parameter $\epsilon$ and the ratio
$r_{\rm in} / R_{\ast}$.
Similarly, the fractional area subtended by both polar caps,
to the north and south of the accretion rings, is given by
\begin{equation}
  \delta_{\rm pol} \, = \, 1 - \cos\theta_{\rm out} \,\, .
\end{equation}
Dupree et al.\  (2005) called this quantity $\phi$ and estimated
it to be $\sim$0.3 for TW Hya.

Accreting gas is assumed to be accelerated from rest at the
inner edge of the disk, and thus is flowing at roughly the ballistic
free-fall speed at the stellar surface,
\begin{equation}
  v_{\rm ff} \, = \, \left[ \frac{2 G M_{\ast}}{R_{\ast}}
  \left( 1 - \frac{R_{\ast}}{r_{\rm in}} \right) \right]^{1/2}
  \,\, .
  \label{eq:vff}
\end{equation}
This slightly underestimates the mean velocity, since in reality
the streams come from radii between $r_{\rm in}$ and $r_{\rm out}$.
Using this expression allows the ram pressure at the stellar surface
to be computed, with
\begin{equation}
  P_{\rm ram} \, = \, \frac{\rho v_{\rm ff}^2}{2} \, = \,
  \frac{v_{\rm ff} \dot{M}_{\rm acc}}{8 \pi \delta R_{\ast}^2}
  \label{eq:Pram}
\end{equation}
(e.g., Hartmann et al.\  1997; Calvet \& Gullbring 1998).
The accretion is assumed to be stopped at the point where
$P_{\rm ram}$ is balanced by the stellar atmosphere's gas pressure.
A representative T Tauri model atmosphere (with gray opacity and
local thermodynamic equilibrium) was used to determine the height
dependence of density and gas pressure.
The atmospheric density $\rho_{\rm sh}$ was thus defined as that of
the highest point in the atmosphere that remains undisturbed by either
the shock or its (denser) post-shock cooling zone.
The age dependence of $\rho_{\rm sh}$ was computed from the
condition of ram pressure balance.
Examining these resulting values, though, yielded a useful
approximation for this quantity.
For most of the evolutionary ages considered for the solar-mass
T Tauri star, the accretion is stopped a few scale heights above
the photosphere, at which the temperature has reached its minimum
radiative equilibrium value of $\sim 0.8 T_{\rm eff}$.
This value can be used in the definition of gas pressure to
obtain a satisfactory estimate for $\rho_{\rm sh}$; i.e., one can
solve for
\begin{equation}
  \rho_{\rm sh} \, \approx \,
  \frac{5 m_{\rm H} P_{\rm ram}}{4 k_{\rm B} T_{\rm eff}}
  \label{eq:rhosh}
\end{equation}
using equation (\ref{eq:Pram}) above and obtain a value within
about 10\% to 20\% of the value interpolated from the full stellar
atmosphere model.
All of these values are only order-of-magnitude estimates, of course,
since neither the post-shock heating nor the thermal equilibrium
{\em inside} the accretion columns has been taken into account
consistently.

In addition to the density at the shock, the models below also
require specifying the density at the visible stellar surface.
A representative photospheric mass density $\rho_{\ast}$ was
computed from the criterion that the Rosseland mean optical
depth should have a value of approximately one:
\begin{equation}
  \tau_{\rm R} \, \approx \,
  \kappa_{\rm R} \rho_{\ast} H_{\ast} \, = \, 1
\end{equation}
where $H_{\ast}$ is the photospheric scale height as defined
above and $\kappa_{\rm R}$ is the Rosseland mean opacity
(in cm$^2$ g$^{-1}$) interpolated as a function of temperature
and pressure from the table of Kurucz (1992).
The resulting ratio $\rho_{\rm sh} / \rho_{\ast}$ varies
strongly over the T Tauri evolution, from a value of order unity
(for the youngest stars with the strongest accretion) down to
$10^{-3}$ at the end of the accretion phase.

\begin{figure}
\epsscale{1.05}
\plotone{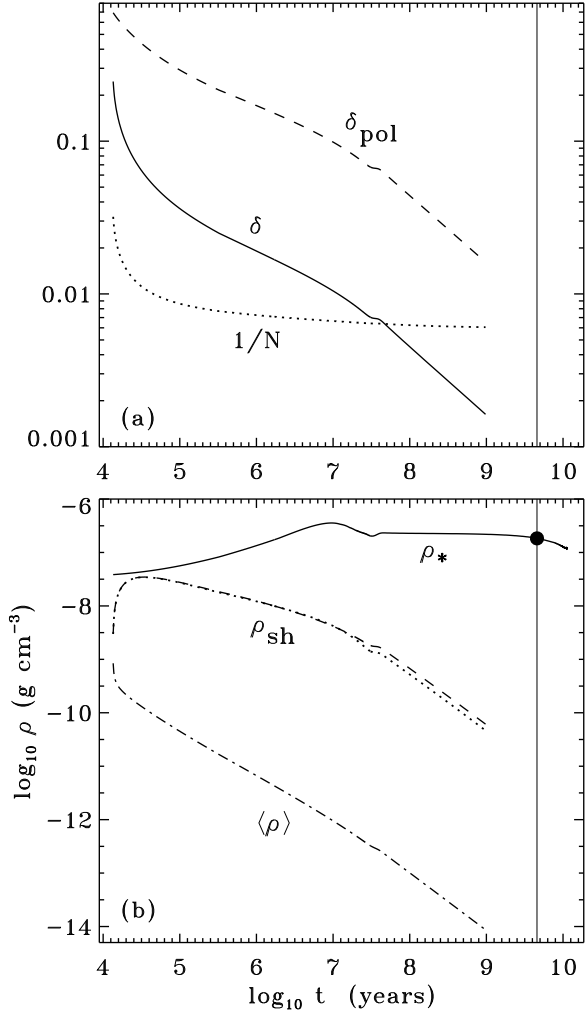}
\caption{Accretion stream properties as a function of age.
({\em{a}}) Filling factor of accretion columns that thread the
disk ({\em solid line}), filling factor of polar caps containing
open field lines ({\em dashed line}), and the reciprocal of the
number of inhomogeneous accretion tubes in one hemisphere
({\em dotted line}).
({\em{b}}) Mass densities at the stellar photosphere
({\em solid line}) and at the accretion shock (exact:
{\em dashed line,} approximation from eq.~[\ref{eq:rhosh}]:
{\em dotted line}), and mean pre-shock density in the accretion
stream ({\em dot-dashed line}).}
\end{figure}

Figure 4 shows several of the accretion-stream quantities defined above.
Figure 4{\em{a}} shows the how the fractional area of both the
accretion streams ($\delta$) and the polar caps ($\delta_{\rm pol}$)
decrease similarly with increasing age.
Calvet \& Gullbring (1998) gave observationally determined values of
$\delta$ for a number of T Tauri stars, but there was no overall
trend with age; $\delta$ was distributed (apparently) randomly
between values of about 0.001 and 0.1 for ages between
$\log_{10} t =4$ and 7.
Dupree et al.\  (2005) estimated $\delta_{\rm pol}$ to be about 0.3,
which agrees with the youngest T Tauri stars modeled here.
Figure 4{\em{b}} plots the age dependence of both the photospheric
density $\rho_{\ast}$ and the density at the accretion shock
$\rho_{\rm sh}$ (computed both numerically and using
eq.~[\ref{eq:rhosh}]).

\subsection{Properties of Inhomogeneous Accretion (``Clumps'')}

Observational evidence for intermittency and time variability
in the accretion streams was discussed in {\S}~2.
Here, the flow along magnetospheric accretion columns is modeled
as an idealized ``string of pearls,'' where infalling dense
clumps (i.e., blobs or clouds) are surrounded by ambient gas of
much lower density.
The clumps are assumed to be roughly spherical in shape with roughly
the same extent in latitude as the magnetic flux tubes connecting
the star and accretion disk.
Thus, once they reach the stellar surface, the clumps have a
radius given by
\begin{equation}
  r_{c} \, = \,
  \frac{R_{\ast} (\theta_{\rm in} - \theta_{\rm out})}{2}
\end{equation}
where the angles $\theta_{\rm in}$ and $\theta_{\rm out}$ must be
expressed in radians.
Figure 2{\em{b}} shows how $r_c$ varies with age.

The dense clumps are assumed to impact the star with a velocity
$v_{c}$ equivalent to the free-fall speed $v_{\rm ff}$ defined in
equation (\ref{eq:vff}).
Scheurwater \& Kuijpers (1988) defined a fiducial shock-crossing time
for clumps impacting a stellar surface, with
$t_{c} = 1.5 r_{c} / v_{c}$.
This time scale is of the same order of magnitude as the time it
would take the blob to pass through the stellar surface if it was
not stopped.
It is useful to assume that the clumps are spread out along the
flux tube with a constant periodicity; i.e., that along a given
flux tube, clumps impact the star at a time interval
$\Delta t = \zeta t_{c}$, where the dimensionless intermittency
factor $\zeta$ must be larger than 1.

When considering the mass accreted by infalling clumps, we follow
Scheurwater \& Kuijpers (1988) and define the density interior
to each clump as $\rho_c$ and the (lower) ambient inter-clump
density as $\rho_{0}$.
For the purpose of mass flux conservation along the magnetic
flux tubes, these densities are defined at the stellar surface.
Despite the fact that the ratio $\rho_{c}/\rho_{0}$ appears in
several resulting quantities below, it always is canceled out
by other quantities that depend separately on $\rho_{c}$ and
$\rho_{0}$ such that the numerical value of their ratio never
needs to be specified directly.
A more relevant quantity for accretion is the mean density
$\langle \rho \rangle$ in the flux tube, which is by definition
larger than $\rho_0$ and smaller than $\rho_c$.
The clump ``overdensity ratio'' ($\rho_{c}/ \langle \rho \rangle$)
can be computed by comparing the volume subtended by one clump
with the volume traversed by a clump over time $\Delta t$.
In other words, if one assumes that $\rho_{c} \gg \rho_{0}$, one
can find the mean density $\langle \rho \rangle$ by spreading
out the gas in each clump to fill its own portion of the flux tube.
Thus,
\begin{equation}
  \frac{\rho_{c}}{\langle \rho \rangle} \, = \,
  \frac{v_{c} \Delta t \, \pi r_{c}^{2}}{4\pi r_{c}^{3} / 3}
  \, = \, \frac{9\zeta}{8} \,\, .
  \label{eq:overden}
\end{equation}
The overdensity ratio (or, equivalently, $\zeta$) is a free
parameter of this system, and a value of
$\rho_{c}/ \langle \rho \rangle = 3$ was chosen more or less
arbitrarily.
This sets $\zeta = 24/9$.
Below, the resulting Alfv\'{e}n wave amplitude (caused by the
periodic clump impacts) is found to depend on an overall factor of
$\zeta^{1/2}$; i.e., it is relatively insensitive to the chosen
value of the overdensity ratio.

In order to relate the inhomogeneous properties along a given
flux tube to the total mass accretion rate $\dot{M}_{\rm acc}$,
the total number of flux tubes impacting the star must be
calculated.
The quantity $N$ is defined as the number of flux tubes
{\em in either the northern or southern hemisphere} that contain
accreting clumps.
This definition is specific to the assumption of an aligned
dipole magnetic field, and it is convenient because the summed
effect of infalling clumps measured at the north [south] pole is
assumed to depend only on the flux tubes in the northern [southern]
hemisphere.
The total number of flux tubes impacting the star is thus assumed
to be $2N$.
One can compute $N$ by comparing two different ways of expressing
the mass accretion rate.
First, we know that in a time-averaged sense, mass is accreted
with a local flux $\langle \rho \rangle v_{c}$ over a subset of
the stellar surface $\delta$ given by equation (\ref{eq:delta}).
Thus,
\begin{equation}
  \dot{M}_{\rm acc} \, = \, 4\pi \delta R_{\ast}^{2}
  \langle \rho \rangle v_{c}  \,\, .
  \label{eq:Mblob1}
\end{equation}
Alternately, the mass in the $2N$ flux tubes can be summed up
by knowing that each clump deposits a mass
$m_{c} = 4\pi r_{c}^{3} \rho_{c} / 3$, with
\begin{equation}
  \dot{M}_{\rm acc} \, = \, \frac{2N m_{c}}{\Delta t} \, = \,
  \frac{16\pi N}{9\zeta} \, \rho_{c} v_{c} r_{c}^{2}  \,\,.
  \label{eq:Mblob2}
\end{equation}
Equations (\ref{eq:Mblob1}) and (\ref{eq:Mblob2}) must give the
same total accretion rate, so equating them gives a useful
expression for
\begin{equation}
  N \, = \, 2 \delta \left( \frac{R_{\ast}}{r_c} \right)^{2}
\end{equation}
where equation (\ref{eq:overden}) was also used.
This quantity is used below to compute the total effect of waves
at the poles from the individual impact events.

Figure 4{\em{a}} shows $N^{-1}$ (rather than $N$ itself, to keep
it on the same scale as the other plotted quantities), and it
is interesting that $N$ remains reasonably constant around
100--150 over most of the classical T Tauri phase of evolution.
This mean value (for any instant of time) is large, but it is not
so large that any fluctuations around the mean would be
unresolvable due to averaging over the star.
If the distribution of flux tubes is assumed to follow some
kind of Poisson-like statistics, a standard deviation of order
$N^{1/2}$ would be expected.
In other words, for $N \approx 100$ there may always be something
like a 10\% level of fluctuations in the magnetospheric accretion
rate.

The equations above also allow $\langle \rho \rangle$ and $\rho_c$
to be computed from the known accretion rate $\dot{M}_{\rm acc}$.
Figure 4{\em{b}} shows $\langle \rho \rangle$ to typically be several
orders of magnitude smaller than both $\rho_{\rm sh}$ and $\rho_{\ast}$.
This large difference arises because the ram pressure inside the
accretion stream is much larger than the gas pressure in the stream.
The ratio $\langle \rho \rangle / \rho_{\rm sh}$ at the stellar
surface is given very roughly by $(c_{s} / v_{c})^2$, where
$c_{s}$ is the sound speed corresponding to $T_{\rm eff}$, and
$c_{s} \ll v_{c}$.
For the youngest T Tauri stars modeled, however,
$c_{s} \approx v_{c}$ and thus 
$\langle \rho \rangle \approx \rho_{\rm sh}$.

\subsection{Properties of Impact-Generated Waves}

This section utilizes the results of Scheurwater \& Kuijpers (1988),
who computed the flux of magnetohydrodynamic (MHD) waves that
arise from the impact of a dense clump onto a stellar surface.
Scheurwater \& Kuijpers (1988) derived the total energy released
in both Alfv\'{e}n and fast-mode MHD waves from such an impact
in the ``far-field'' limit (i.e., at horizontal distances large
compared to the clump size $r_c$).
Slow-mode MHD waves were not considered because of the use of
the cold-plasma approximation.
The models below utilize only their Alfv\'{e}n wave result, since
the efficiency of fast-mode wave generation was found to be
significantly lower than for Alfv\'{e}n waves.
Also, due to their compressibility, the magnetosonic (fast and slow
mode) MHD waves are expected to dissipate more rapidly than
Alfv\'{e}n waves in stellar atmospheres (e.g., Kuperus et al.\  1981;
Narain \& Ulmschneider 1990, 1996; Whang 1997).
Thus, even if fast-mode and Alfv\'{e}n waves were generated in equal
amounts at the impact site, the fast-mode waves may not survive
their journey to the polar wind-generation regions of the star
without being strongly damped.

For simplicity, we assume that waves propagate away from the
impact site with an overall energy given by the
Scheurwater \& Kuijpers (1988) Alfv\'{e}n wave result, and that
the waves undergo negligible dissipation.
The wave energy released in one impact event is given by
\begin{equation}
  E_{\rm A} \, = \, 0.06615 \,\, \frac{\rho_c}{\rho_0}
  \left( \frac{v_c}{V_{\rm A}} \right)^{3} m_{c} v_{c}^{2}
  \label{eq:EA}
\end{equation}
where the numerical factor in front was given approximately as 0.07
in equation (57) of Scheurwater \& Kuijpers (1988), but has been
calculated more precisely from their Bessel-function integral.
In this context, the Alfv\'{e}n speed $V_{\rm A}$ is defined as that
of the ambient medium, with
\begin{equation}
  V_{\rm A} \, = \, \frac{B_{0}}{\sqrt{4\pi\rho_0}}
\end{equation}
and $B_{0} \approx 1000$ G is the ambient magnetic field strength at
the stellar surface, where the accretion streams connect with the star.
Scheurwater \& Kuijpers (1988) assumed that $v_{c} < V_{\rm A}$
(i.e., that the impacting clumps do not strongly distort the
background magnetic field).
This tends to lead to a very low efficiency of wave generation,
such that equation (\ref{eq:EA}) may be considered a ``conservative''
lower limit to the available energy of fluctuations.
For a given accretion column, the wave power that is emitted
continuously by a stream of clumps is given by $E_{\rm A}/ \Delta t$.

In order to compute the wave energy density at other points on the
stellar surface, we make the assumption that the waves propagate
out {\em horizontally} from the impact point.
It is important to note that ideal Alfv\'{e}n waves do not
propagate any energy perpendicularly to the background magnetic field.
There are several reasons, however, why this does not disqualify the
adopted treatment of wave energy propagation over the horizontal
stellar surface.
First, the true evolution of the impact pulse is probably nonlinear.
Much like the solar Moreton and EIT waves discussed in {\S}~2,
nonlinear effects such as mode coupling, shock steepening, and
soliton-like coherence are likely to be acting to help convey the
total ``fluctuation yield'' of the impact across the stellar surface.
Second, for any real star, the background magnetic field is never
completely radial, and it will always have some component horizontal
to the surface (along which even linear Alfv\'{e}n waves can propagate).
Thus, the dominant end-result of multiple cloud impacts is assumed
here to be {\em some} kind of transverse field-line perturbations that
are treated for simplicity with the energetics of Alfv\'{e}n waves.

Considering waves that spread out in circular ripples from
the impact point, the goal is to compute the horizontal wave flux
(power per unit area) at a distance $x$ away from the central point.
For this purpose, the stellar surface can be treated as a flat plane.
The wave power is assumed to be emitted into an expanding cylinder
with an approximate height of $2 r_c$ (the clump diameter) and a
horizontal radius $x$.
The horizontal flux $F_{\rm A}$ of waves into the vertical
wall of the cylinder is given by dividing the power by the
area of the surrounding wall, with
\begin{equation}
  F_{\rm A} \, = \, \frac{E_{\rm A} / \Delta t}{4\pi x r_{c}}
  \, = \, 0.0147 \,\,
  \frac{r_{c} \rho_{c}^{2} v_{c}^{3}}{x \rho_{0} \zeta}
  \left( \frac{v_c}{V_{\rm A}} \right)^{3} \,\, .
\end{equation}
Note that the flux decreases linearly with increasing $x$, as is
expected for a cylindrical geometry.

The total wave flux from the effect of multiple clump impacts is
computed at the north or south pole of the star.
The accretion stream impact points are assumed to be distributed
circularly in a ring around the pole.
Thus, using the geometric quantities derived earlier, this implies
$x = R_{\ast} (\theta_{\rm in} + \theta_{\rm out})/2$.
Also, because each accretion column is assumed to be at the same
horizontal distance from the target point at the pole, the total
wave flux is given straightforwardly by $N$ times the flux due
to one stream of clumps.
The total accretion-driven wave flux arriving at either the north
or south pole is thus given by $F_{\rm A, tot} = N F_{\rm A}$.
It is assumed that the waves do not damp appreciably between where
they are generated (at the bases of the accretion streams) and their
destination (at the pole).

The standard definition of the flux of Alfv\'{e}n waves, in a
medium where the bulk flow speed is negligible in comparison to
the Alfv\'{e}n speed, is
\begin{equation}
  F_{\rm A, tot} \, = \, \rho_{0} v_{\perp}^{2} V_{\rm A}
  \,\, .
\end{equation}
This can be compared with the total wave flux derived above to
obtain the perpendicular Alfv\'{e}n wave velocity amplitude
$v_{\perp}$.
In units of the clump infall velocity $v_c$, the wave amplitude
is thus given by
\begin{equation}
  \frac{v_{\perp}}{v_c} \, = \, 0.1715 \,\, \frac{\rho_c}{\rho_0} 
  \left( \frac{N r_c}{\zeta x} \right)^{1/2}
  \left( \frac{v_c}{V_{\rm A}} \right)^{2} \,\, .
  \label{eq:vperpsh}
\end{equation}
Note that there is no actual dependence on $\rho_0$, since
the explicit factor of $\rho_0$ in the denominator is canceled
by the density dependence of $V_{\rm A}^{-2}$.

The wave amplitude derived in equation (\ref{eq:vperpsh}) is the
value at the shock impact height, which formally can be above
or below the photosphere.
The stellar wind models below, though, require the Alfv\'{e}n wave
amplitude to be specified exactly at the photosphere.
If we assume that the wave energy density
$U_{\rm A} = \rho v_{\perp}^{2}$ is conserved between the
shock height and the photosphere, then the densities at those
heights can be used to scale one to the other, with
\begin{equation}
  v_{\perp \ast} \, = \, v_{\perp}
  \left( \frac{\rho_{\rm sh}}{\rho_{\ast}} \right)^{1/2}
  \label{eq:vpast}
\end{equation}
where $v_{\perp\ast}$ is the photospheric wave amplitude, and
the densities $\rho_{\ast}$ and $\rho_{\rm sh}$ were defined
above in {\S}~3.2.

Although the overall energy budget of accretion-driven waves is
treated under the assumption that the waves are Alfv\'{e}nic,
it is likely that the strongly nonlinear stream impacts also give
rise to compressible waves of some kind.
As mentioned above, the analysis of Scheurwater \& Kuijpers (1988)
did not take account of slow-mode MHD waves that, for parallel
propagation and a strong background field, are identical to
hydrodynamic acoustic waves.
There is evidence, however, that another highly nonlinear
MHD phenomenon---turbulent subsurface convection---gives
rise to both longitudinal (compressible) and transverse
(incompressible) MHD waves with roughly comparable energy
densities (e.g., Musielak \& Ulmschneider 2002).
Thus, the models below are given a photospheric source of
accretion-driven {\em acoustic waves} that are in energy
equipartition with the accretion-driven Alfv\'{e}n waves; i.e.,
$U_{s} = U_{\rm A}$.
The upward flux of acoustic waves is thus given by
$F_{s} = c_{s} U_{s}$, where $c_{s}$ is the sound
speed appropriate for $T_{\rm eff}$.

For both the acoustic and Alfv\'{e}n waves at the photosphere,
the accretion-driven component is added to the intrinsic
(convection-driven) component.
A key assumption of this paper is that the convection-driven
component is held fixed, as a function of age, at the values
used by Cranmer et al.\  (2007) for the present-day Sun.
This results in minimum values for the Alfv\'{e}n wave amplitude
(0.255 km s$^{-1}$) and the acoustic wave flux
($10^8$ erg cm$^{-2}$ s$^{-1}$) below which the models never go.
There are hints that rapidly rotating young stars may undergo
more intense subsurface convection than the evolved slowly
rotating Sun (K\"{a}pyl\"{a} et al.\  2007; Brown et al. 2007;
Ballot et al. 2007), but the implications of these additional
variations with age are left for future work (see also {\S}~6).

\begin{figure}
\epsscale{1.08}
\plotone{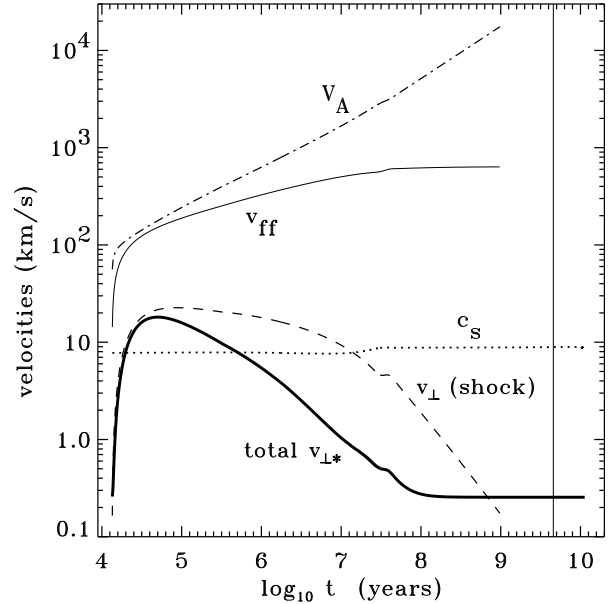}
\caption{Velocities related to accretion streams as a function
of age: free-fall clump speed ({\em thin solid line}), ambient
Alfv\'{e}n speed ({\em dot-dashed line}), photospheric sound
speed ({\em dotted line}).
Plotted Alfv\'{e}n wave amplitudes are those due to
accretion-streams and measured at the shock
({\em dashed line}) and those due to both accretion and
convection, measured at the photosphere ({\em thick solid line}).}
\end{figure}

Figure 5 displays various velocity quantities used in the
accretion-driven wave scenario.
The clump free-fall speed $v_{c} = v_{\rm ff}$ always remains
smaller than the ambient Alfv\'{e}n speed $V_{\rm A}$, which
was an assumption that Scheurwater \& Kuijpers (1988) had to make
in order for the background magnetic field to remain relatively
undisturbed by the clumps.
The accretion-driven Alfv\'{e}n wave amplitude at the shock is
larger than that measured at the photosphere (see
eq.~[\ref{eq:vpast}]), and at ages later than about 60 Myr
the accretion-driven waves at the photosphere grow weak enough
to be overwhelmed by the convection-driven waves.
For a limited range of younger ages
($2 \times 10^{4} < t < 5 \times 10^5$ yr) the Alfv\'{e}n wave
motions are supersonic in the photosphere, with a peak value
of $v_{\perp\ast} = 18$ km s$^{-1}$ at $t = 5 \times 10^{4}$ yr.
The sharp decrease in wave amplitude at the youngest ages is
due to the inner edge of the accretion disk coming closer to
the stellar surface.
In that limit, the ballistic infall speed $v_{\rm ff}$
becomes small and the latitudinal distance $x$ traversed by
the waves becomes large, thus leaving negligible energy in
the waves once they reach the pole.

Finally, it is worthwhile to sum up how the above values depend
on the relatively large number of assumptions made about the
accretion streams.
First, the use of an aligned dipole field for the accretion
streams, the K\"{o}nigl (1991) expression for $r_{\rm in}$,
and the ``string of pearls'' model of clumps along the flux
tubes are all somewhat simplistic and should be replaced by
more realistic conditions in future work.
Second, there are three primary parameters that had to be
specified in order to determine numerical values for the
accretion properties.  These are:
(1) the relative size of the outer disk radius with respect to
the inner disk radius; i.e., $\epsilon = 0.1$, (2) the clump
overdensity ratio $\rho_{c} / \langle\rho\rangle = 3$, and
(3) the magnetic field strength at the base of the accretion
streams $B_{0} = 1000$ G.
Third, probably the most idealistic assumption in the modeled
evolutionary sequence is the use of a single monotonic relation
for $\dot{M}_{\rm acc}$ versus $t$ (e.g., eq.~[\ref{eq:Maccfit}]).
{\S}~6 describes how these assumptions can be relaxed in
subsequent modeling efforts.

\section{Implementation in Stellar Wind Models}

The steady-state outflow models presented here are numerical
solutions to one-fluid conservation equations for mass, momentum,
and energy along a polar flux tube.
For the specific case of the solar wind, Cranmer et al.\  (2007)
presented these equations and described their self-consistent
numerical solution using a computer code called ZEPHYR.
The T Tauri wind models are calculated with an updated version of
ZEPHYR, with specific differences from the solar case described
below.

\subsection{Conservation Equations and Input Physics}

The equation of mass conservation along a magnetic flux tube is
\begin{equation}
  \frac{1}{A} \frac{\partial}{\partial r} \left( \rho u A \right)
  \, = \, 0
  \label{eq:drhodt}
\end{equation}
where $u$ and $\rho$ are the wind speed and mass density specified
as functions of radial distance $r$, and $A$ is the cross-sectional
area of the polar flux tube.
Magnetic flux conservation demands that the product $B_{0}A$ is
constant along the flux tube, where $B_{0}(r)$ is the field strength
that is specified explicitly.

The equation of momentum conservation is
\begin{equation}
  u \frac{\partial u}{\partial r}
  + \frac{1}{\rho} \frac{\partial P}{\partial r} \, = \,
  - \frac{GM_{\ast}}{r^2} + D
\end{equation}
where $P$ is the gas pressure and $D$ is the bulk acceleration
on the plasma due to {\em wave pressure;} i.e., the nondissipative
net ponderomotive force due to wave propagation through the
inhomogeneous medium (Bretherton \& Garrett 1968; Belcher 1971;
Jacques 1977).
A complete expression for $D$ in the presence of damped acoustic
and Alfv\'{e}n waves was given by Cranmer et al.\  (2007).
The simpler limit of wave pressure due to dissipationless
Alfv\'{e}n waves is discussed in more detail in {\S\S}~5.2--5.3.

For the pure hydrogen plasma assumed here, the gas pressure is
given by $P = (1 + x) \rho k_{\rm B} T / m_{\rm H}$, where $x$ is
the hydrogen ionization fraction.
Note that although the pressure is calculated for a hydrogen gas,
the radiative cooling rate $Q_{\rm rad}$ used in the energy equation 
is dominated by metals.\footnote{%
Although this is formally inconsistent, the resulting properties of
the plasma are not expected to be far from values computed with a
more accurate equation of state.}
Cranmer et al.\  (2007) tested the ZEPHYR code with two separate
assumptions for the ionization balance.
First, a self-consistent, but computationally intensive solution
was used, which implemented a three-level hydrogen atom.
In that model, the $n=1$ and $n=2$ levels were assumed to remain in
relative local thermodynamic equilibrium (LTE) and the full rate
equation between $n=2$ and the continuum was solved iteratively.
Second, a simpler tabulated function of $x$ as a function of
temperature $T$ was taken from a semi-empirical non-LTE model
of the solar photosphere, chromosphere, and transition region
(e.g., Avrett \& Loeser 2008).
For both the solar and T Tauri star applications, the results for
the two cases were extremely similar, and thus the simpler
tabulated function was used in the models described below.

For solar-type winds, the key equation for both heating the corona
and setting the mass loss rate is the conservation of internal energy,
\begin{equation}
  u \frac{\partial E}{\partial r}
  + \left( \frac{E+P}{A} \right)
  \frac{\partial}{\partial r} \left( u A \right)
  \, = \, Q_{\rm rad} + Q_{\rm cond} + Q_{\rm A} + Q_{\rm S}
  \label{eq:dEdt}
\end{equation}
where $E$ is the internal energy density and the terms on the
right-hand side are volumetric heating/cooling rates due to
radiation, conduction, Alfv\'{e}n wave damping, and acoustic (sound)
wave damping.
The terms on the left-hand side that depend on $u$ are responsible
for enthalpy transport and adiabatic cooling.
In a partially ionized plasma, the definition of $E$ is
convention-dependent; we use the same one as
Ulmschneider \& Muchmore (1986) and Mullan \& Cheng (1993),
which is $E = (3P/2) + x \rho I_{\rm H} / m_{\rm H}$, where
$I_{\rm H} = 13.6$ eV.
The net heating/cooling from conduction ($Q_{\rm cond}$) is given
by a gradual transition between classical Spitzer-H\"{a}rm
conductivity (when the electron collision rate is fast compared
to the wind expansion rate) and Hollweg's (1974, 1976) prescription
for free-streaming heat flux in the collisionless heliosphere
(when electron collisions are negligible).
All of these terms were described in more detail by
Cranmer et al.\  (2007).

The volumetric radiative heating/cooling rate $Q_{\rm rad}$ has
been modified slightly from the earlier solar models.
The solar rate is first computed as in Cranmer et al.\  (2007),
but then it is multiplied by a correction factor similar to that
suggested by Hartmann \& Macgregor (1980) for massive,
{\em optically thick} chromospheres of late-type stars.
The correction factor $f$ is assumed to be proportional to the
escape probability $P_{\rm esc}$ for photons in the core of
the \ion{Mg}{2} $\lambda$2803 ($h$) and $\lambda$2796 ($k$) lines.
A simple expression that bridges the optically thin and thick
limits, for a line with Voigt wings, is
\begin{equation}
  f \, = \, 2 P_{\rm esc} \, \approx \,
  \frac{1}{1 + \tau_{\rm hk}^{1/2}}
\end{equation}
(e.g., Mihalas 1978), which may err on the side of overestimating
the escape probability and thus would give a conservative
undercorrection to $Q_{\rm rad}$.
Hartmann \& Macgregor (1980) assumed that the optical depth of
low-ionization metals would scale as $g^{-1/2}$, where $g$ is the
stellar surface gravity.
Here, we compute the optical depth in the core of the $h$ and $k$
lines ($\tau_{\rm hk}$) more exactly by integrating over the radial
grid of density and temperature in the wind model,
\begin{equation}
  \tau_{\rm hk} \, = \, \int \chi_{\rm hk}
  \left( \frac{R_{\ast}}{r} \right)^{2} dr  \,\, ,
\end{equation}
where we include the spherical correction factor suggested by
Lucy (1971, 1976) for extended atmospheres.
The line-center extinction coefficient is given approximately by
\begin{equation}
  \chi_{\rm hk} \, = \, 0.0153 \,
  \left( \frac{n_{\rm Mg \, II}}{10^{10} \,\, \mbox{cm}^{-3}} \right)
  \left( \frac{T}{10^{4} \,\, \mbox{K}} \right)^{-1/2}
  \, \mbox{cm}^{-1}
\end{equation} 
and the number density $n_{\rm Mg \, II}$ of ground-state ions
is given by the product of the hydrogen number density
($\rho / m_{\rm H}$), the Mg abundance with respect to hydrogen
($3.4 \times 10^{-5}$; Grevesse et al.\  2007), and the ionization
fraction of Mg$^{+}$ with respect to all Mg species.
For temperatures in excess of $10^4$ K, the latter is given by
coronal ionization equilibrium (Mazzotta et al.\  1998); for
temperatures below $10^4$ K, the coronal equilibrium curve is
matched onto a relation determined from LTE Saha ionization
balance.

Following Hartmann \& Macgregor (1980), the opacity correction
factor $f$ is ramped up gradually as a function of temperature.
When $T < 7160$ K, $Q_{\rm rad}$ is multiplied by the full
factor $f$.
For $7160 < T < 8440$ K, the multiplier is varied by taking
$f^{(8440-T)/1280}$, which ramps down to unity by the time the
temperature rises to 8440 K.
For temperatures in excess of 8440 K, the correction factor is
not used.

The evolution and damping of Alfv\'{e}n and acoustic waves---which
affects both $D$ in the momentum equation and $Q_{\rm A}$ and
$Q_{\rm S}$ in the energy equation---is described in much more
detail by Cranmer et al.\  (2007).
Given a specified frequency spectrum of acoustic and Alfv\'{e}n wave
power in the photosphere, equations of wave action conservation are
solved to determine the radial evolution of the wave amplitudes.
Acoustic waves are damped by both heat conduction and entropy gain
at shock discontinuities.
Alfv\'{e}n waves are damped by MHD turbulence, for which we only
specify the net transport of energy from large to small eddies,
assuming the cascade must terminate in an irreversible conversion
of wave energy to heat.

Coupled with the wave action equations are also non-WKB transport
equations to determine the degree of linear reflection of the
Alfv\'{e}n waves (e.g., Heinemann \& Olbert 1980).
This is required because the turbulent dissipation rate depends on
differences in energy density between upward and downward
traveling waves (see also Matthaeus et al.\  1999;
Dmitruk et al.\  2001, 2002; Cranmer \& van Ballegooijen 2005).
The resulting values of the Elsasser amplitudes $Z_{\pm}$, which
denote the energy contained in upward ($Z_{-}$) and downward
($Z_{+}$) propagating waves, were then used to constrain the
energy flux in the cascade.
The Alfv\'{e}n wave heating rate (erg s$^{-1}$ cm$^{-3}$) is
given by a phenomenological relation that has evolved from analytic
studies and numerical simulations; i.e.,
\begin{equation}
  Q_{\rm A} \, = \, \rho \, \left(
  \frac{1}{1 + t_{\rm eddy}/t_{\rm ref}} \right) \,
  \frac{Z_{-}^{2} Z_{+} + Z_{+}^{2} Z_{-}}{4 \ell_{\perp}}
  \label{eq:Qdmit}
\end{equation}
(see also Hossain et al.\  1995; Zhou \& Matthaeus 1990;
Oughton et al.\  2006).
The transverse length scale $\ell_{\perp}$ is an effective
perpendicular correlation length of the turbulence, and
Cranmer et al.\  (2007) used a standard assumption that
$\ell_{\perp}$ scales with the cross-sectional width of the
flux tube (Hollweg 1986).
The term in parentheses in equation (\ref{eq:Qdmit}) is an
efficiency factor that accounts for situations in which the 
turbulent cascade does not have time to develop before the waves
or the wind carry away the energy (Dmitruk \& Matthaeus 2003).
The cascade is quenched when the nonlinear eddy time scale
$t_{\rm eddy}$ becomes much longer than the macroscopic wave
reflection time scale $t_{\rm ref}$.

\subsection{Model Inputs and Numerical Procedures}

The ZEPHYR code was designed to utilize as few free parameters
as possible.
For example, the coronal heating rate and the spatial length
scales of wave dissipation are computed {\em internally} from
straightforward physical principles and are not input as
adjustable parameters.
However, there are quantities that do need to be specified
prior to solving the equations given in {\S}~4.1.

An important input parameter to ZEPHYR is the radial dependence
of the background magnetic field $B_{0}(r)$.
Although many stellar field-strength measurements have been made,
relatively little is known about how rapidly $B_0$ decreases
with increasing height above the photosphere, or how fragmented
the flux tubes become on the granulation scale.
Some information about this kind of structure is contained in
``filling factors'' that can be determined observationally
(e.g., Saar 2001).
However, these measurements may be biased toward the bright
closed-field active regions and not the footpoints of stellar
wind streams.
Thus, as in other areas of this study, the present-day solar case
was adopted as a baseline on which to vary the evolving properties
of the 1 $M_{\odot}$ model star.

\begin{figure}
\epsscale{1.08}
\plotone{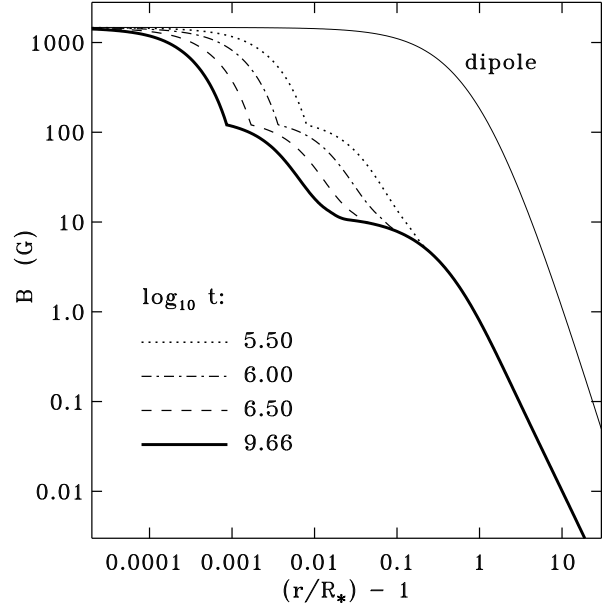}
\caption{Background magnetic field $B_{0}$ over the stellar poles as a
function of the height above the photosphere (measured in
stellar radii).
Present-day solar field strength ({\em thick solid line}) is
compared with that of ages: $\log_{10} t = 5.5$ ({\em dotted line}),
6.0 ({\em dot-dashed line}), and 6.5 ({\em dashed line}).
Also shown is an ideal dipolar field having the same strength
at the photosphere as the other cases ({\em thin solid line}).}
\end{figure}

The multi-scale polar coronal hole field of
Cranmer \& van Ballegooijen (2005) was used as the starting
point for the younger T Tauri models.
The lower regions of that model were derived for a magnetostatic
balance between the strong-field (low density) solar wind flux
tube and the weak-field (high density) surrounding atmosphere.
The radial dependence of magnetic pressure $B_{0}^{2}/8\pi$
thus scales with the gas pressure $P$.
In the T Tauri models, then, the lower part of the magnetic field
was stretched in radius in proportion with the pressure scale
height $H_{\ast}$.
Figure 6 shows $B_{0}(r)$ for several ages, along with an
idealized dipole that does not take account of the lateral
expansion of the flux tube close to the surface.

In addition to the global magnetic field strength, the ZEPHYR
models also require three key wave-driving parameters to be
specified at the lower boundary:
\begin{enumerate}
\item
The {\em photospheric acoustic flux} $F_{s}$ mainly affects
the heating at chromospheric temperatures ($T \sim 10^{4}$ K).
The solar value of $10^8$ erg cm$^{-2}$ s$^{-1}$ was summed with
the accretion-driven acoustic flux as discussed in {\S}~3.4.
The power spectrum of acoustic waves was adopted from the solar
spectrum given in Figure 3 of Cranmer et al.\  (2007), and the
frequency scale was shifted up or down with the photospheric value
of the acoustic cutoff frequency,
$\omega_{\rm ac} = c_{s} / 2H_{\ast}$, which evolves over time.
\item
The {\em photospheric Alfv\'{e}n wave amplitude} $v_{\perp \ast}$
is specified instead of the upward energy flux $F_{\rm A}$ because
the latter depends on the cancellation between upward and downward
propagating waves that is determined as a part of the
self-consistent solution.
As discussed in {\S}~3.4, the solar value of 0.255 km s$^{-1}$
was supplemented by the accretion-driven wave component.
The shape of the Alfv\'{e}n wave frequency spectrum was kept fixed
at the solar model because it is unclear how it should scale with
varying stellar properties.
\item
The {\em photospheric Alfv\'{e}n wave correlation length}
$\ell_{\perp \ast}$ sets the scale of the turbulent heating rate
$Q_{\rm A}$ (eq.\  [\ref{eq:Qdmit}]).
Once this parameter is set, the value of $\ell_{\perp}$ at larger
distances is determined by the assumed proportionality with
$A^{1/2}$.
The solar value of 75 km (Cranmer et al.\  2007) was evolved in
proportion with changes in the photospheric scale height $H_{\ast}$.
The justification for this is that the horizontal scale 
of convective granulation is believed to be set by the scale height
(e.g., Robinson et al.\  2004), and the turbulent mixing scale is
probably related closely to the properties of the granulation.
\end{enumerate}

The numerical relaxation method used by ZEPHYR was discussed by
Cranmer et al.\  (2007).
In the absence of explicit specification here, the new T Tauri
wind models use the same parameters as given in that paper.
However, the new models use a slightly stronger form for the code's
iterative undercorrection than did the original solar models.
From one iteration to the next, ZEPHYR replaces old values with
with a fractional step toward the newly computed values, rather
than using the new values themselves.
This technique was motivated by globally convergent backtracking
methods for finding roots of nonlinear equations (e.g.,
Dennis \& Schnabel 1983).
The solar models used a constant minimum undercorrection exponent
$\epsilon_{0} = 0.17$, as defined in equation (65) of
Cranmer et al.\  (2007).
The T Tauri models started with this value, but gradually decreased
it over time by multiplying $\epsilon_0$ by 0.97 after each iteration.
The value of $\epsilon_{0}$ was not allowed to be smaller than
an absolute minimum of 0.001.
This represented an additional kind of ``annealing'' that helped
the parameters reach their time-steady values more rapidly
and robustly.

\section{Results}

\subsection{Standard Age Sequence}

\begin{figure}
\epsscale{1.05}
\plotone{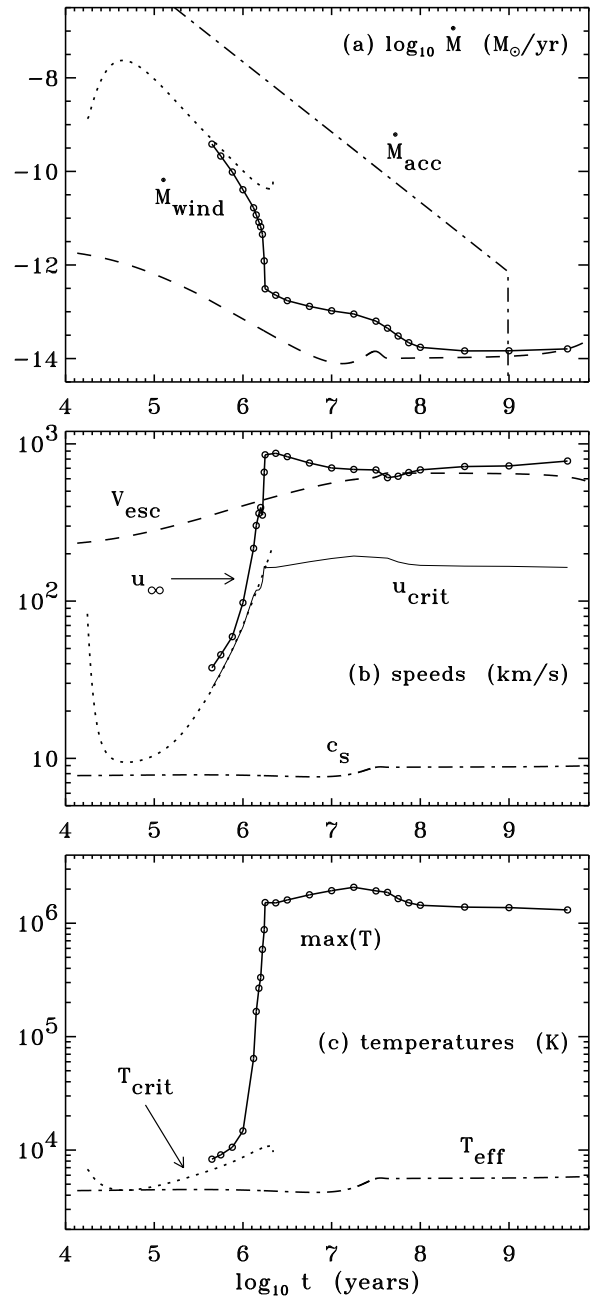}
\caption{Age-dependent model properties.
({\em{a}}) Mass loss rates of time-steady ZEPHYR models
({\em solid line with symbols}) compared with the Reimers (1975, 1977)
relation ({\em dashed line}), analytic/cold models of {\S}~5.3
({\em dotted line}), and adopted accretion rate
({\em dot-dashed line}).
({\em{b}}) Terminal wind speed for ZEPHYR models
({\em thick solid line with symbols}), surface escape speed
({\em dashed line}), photospheric sound speed ({\em dot-dashed line}),
and wind speed at the wave-modified critical point for ZEPHYR
models ({\em thin solid line}) and analytic/cold models
({\em dotted line}).
({\em{c}}) Peak temperatures of ZEPHYR models
({\em solid line with symbols}), temperatures at critical point
for analytic/cold models ({\em dotted line}), and stellar
$T_{\rm eff}$ ({\em dot-dashed line}).}
\end{figure}

A series of 24 ZEPHYR models was created with ages ranging
between $\log_{10} t = 5.65$ and 9.66.
These models all converged to steady-state mass-conserving wind
outflow solutions within 250 numerical iterations.
The relative errors in the energy equation, averaged over the radial
grid, were of order 1\% for the final converged models.
For ages younger than $\log_{10} t = 5.65$ (i.e., 0.45 Myr),
it was found that time-steady solutions to
equations (\ref{eq:drhodt})--(\ref{eq:dEdt}) do not exist, and the
best we can do is to estimate the mass flux that is driven up through
the Parker critical point (see {\S\S}~5.2--5.3 below).
Figure 7 presents a summary of various scalar properties for the
wind models as a function of age.

The mass loss rate $\dot{M}_{\rm wind}$, shown in Figure 7{\em{a}},
was calculated by multiplying the mass flux $\rho u$ at the largest
grid radius ($1200 \, R_{\ast}$ for all models) by the full spherical
area $4\pi r^{2}$ at that radius.
This is a slight overestimate, since the polar flux tubes only cover
a finite fraction of the stellar surface ($\delta_{\rm pol} < 1$).
However, it is unknown whether these regions expand outward or inward
(i.e., to larger or smaller solid angle coverage) with increasing
distance from the star.
The actual large-scale geometry is likely to depend on how the pressure
(gas and magnetic) in the field lines that thread the accretion disk
is able to confine or collimate the polar flux tubes.

As the age is decreased, $\dot{M}_{\rm wind}$ for the time-steady
models increases by four orders of magnitude, from the present-day solar
value of about $2 \times 10^{-14}$ $M_{\odot}$ yr$^{-1}$ up to 
$4 \times 10^{-10}$ $M_{\odot}$ yr$^{-1}$ at the youngest modeled
age of 0.45 Myr.
Note that these values exceed the mass loss rates predicted by the
empirical scaling relation of Reimers (1975, 1977) at all ages (i.e.,
using $\dot{M}_{\rm wind} \propto L_{\ast} R_{\ast} / M_{\ast}$ and
normalizing to the present-day mass loss rate), but for
$t\gtrsim 100$ Myr there is rough agreement.
The stellar wind velocity at the largest radius is denoted as
an ``asymptotic'' or terminal speed $u_{\infty}$ and is shown
in Figure 7{\em{b}}.
The wind speed remains roughly constant for most of the later phase
of the evolution (with $u_{\infty} \approx V_{\rm esc}$), but it
drops precipitously in the youngest models.

The dominant physical processes that drive the evolving stellar winds
are revealed when examining the maximum temperatures in the models,
as shown in Figure 7{\em{c}}.
The older ($t \gtrsim 1.75$ Myr) models with high wind speeds and
solar-like mass loss rates have hot coronae, with peak temperatures
between $10^6$ and $2 \times 10^6$ K.
The younger models undergo a rapid drop in temperature, ultimately
leading to ``extended chromospheres'' with peak temperatures
around $10^4$ K.
This transition occurs because of a well-known {\em thermal
instability} in the radiative cooling rate $Q_{\rm rad}$
(e.g., Parker 1953; Field 1965; Rosner et al.\  1978; Suzuki 2007).
At temperatures below about $10^5$ K, the cooling rate decreases
with decreasing $T$, and small temperature perturbations are
easily stabilized.
However, above $10^5$ K, $Q_{\rm rad}$ decreases as $T$ increases,
which gives any small increase in temperature an unstable
runaway toward larger values.
This is the same instability that helps lead to the sharp transition
region between the $10^4$ K solar chromosphere and the $10^6$ K corona
(see also Hammer 1982; Withbroe 1988; Owocki 2004).

The reason that the young and old models in Figure 7 end up on
opposite sides of the thermal instability is that the total rates
of energy input (i.e., the wave heating rates $Q_{\rm A}$ and
$Q_{\rm S}$) vary strongly as a function of age.
The older models have lower wave amplitudes and thus weaker
heating rates, which leads to relatively low values of
$\dot{M}_{\rm wind}$.
The correspondingly low atmospheric densities in these models give
rise to weak radiative cooling (because $Q_{\rm rad} \propto \rho^{2}$)
that cannot suppress the coronal heating.
The coronal winds are driven by comparable contributions from gas
pressure and wave pressure.
On the other hand, the younger models have larger wave amplitudes,
more energy input (as well as more wave pressure, which expands the
density scale height), and thus more massive winds with
stronger radiative cooling.
These chromospheric winds are driven mainly by wave pressure.
Related explanations for cool winds from low-gravity stars
have been discussed by, e.g., Antiochos et al.\  (1986),
Rosner et al.\  (1991, 1995), Willson (2000), Killie (2002), and
Suzuki (2007).

The processes that determine the quantitative value of
$\dot{M}_{\rm wind}$ are different on either side of the
thermal bifurcation.
The mass flux for the older, solar-like models can be explained well
by the energy balance between heat conduction, radiative losses,
and the upward enthalpy flux.
The atmosphere finds the height of the transition region that
best matches these sources and sinks of energy in a time-steady way,
and the resulting gas pressure at this height sets the mass flux.
Various analytic solutions for this balance have been given by
Hammer (1982), Withbroe (1988), Leer et al.\  (1998), and
Cranmer (2004).
The younger, more massive wind models are found to approach the
limit of ``cold'' wave-driven outflows, where the Alfv\'{e}n wave
pressure replaces the gas pressure as the primary means of
canceling out the stellar gravity.
In this limit, the analytic approach of Holzer et al.\  (1983),
discussed further in {\S}~5.3, has been shown to provide a
relatively simple estimate for the mass loss rate.

\begin{figure}
\epsscale{1.05}
\plotone{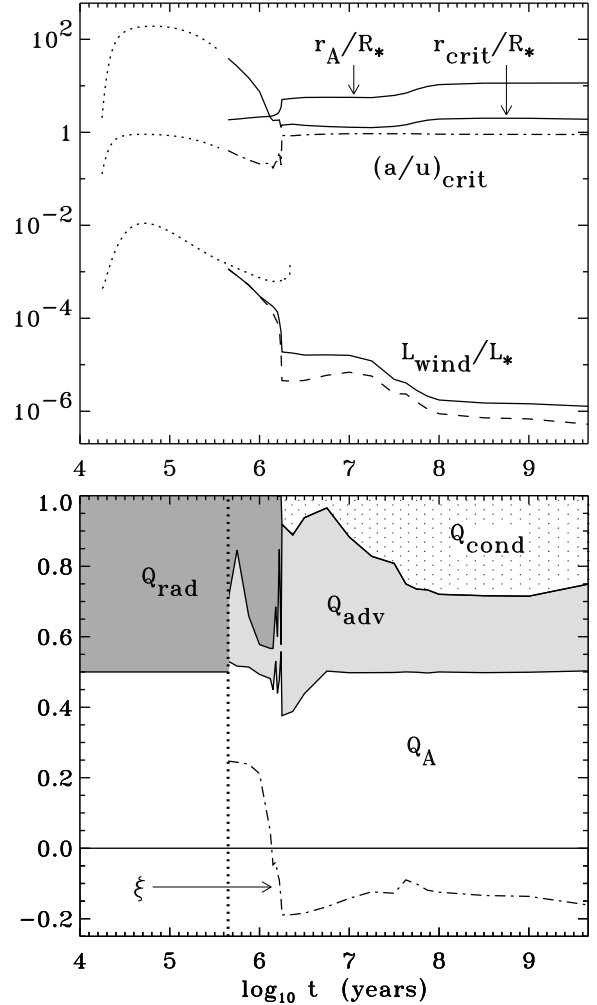}
\caption{Additional age-dependent model properties.
({\em{a}}) Wave-modified critical radii and Alfv\'{e}n radii
({\em upper solid lines, labeled}), ratio of sound speed to
wind speed at critical radius ({\em dot-dashed line}), and
ratio of wind luminosity to photon luminosity, measured at
the largest modeled radius ({\em lower solid line}) and at
the critical point ({\em dashed line}).
Results for the analytic/cold models are also shown
({\em dotted lines}).
({\em{b}}) Areas denote the relative contribution of terms to the
energy conservation equation at the critical point; see labels.
($Q_{\rm adv}$ denotes the advection terms on the left-hand
side of equation (\ref{eq:dEdt}), other terms are defined
in the text.)  Also plotted is the exponent $\xi$ versus age
({\em dot-dashed line}).}
\end{figure}

Figure 8 illustrates a selection of other parameters of the ZEPHYR
models.
The heights of the wave-modified critical point (see
eq.~[\ref{eq:ucrit}] below) and the Alfv\'{e}n point (where
$u = V_{\rm A}$) are shown in units of the stellar radius.
The ratio of the isothermal sound speed $a$ to the wind speed at
the critical point is also shown.
This ratio is close to unity for the older, less massive models
(indicating that gas pressure dominates the wind acceleration) and
is less than 1 for the younger (more wave-driven) models.
Note also that the older ``coronal'' models have critical points in
the sub-Alfv\'{e}nic wind ($u < V_{\rm A}$) and the younger
``extended chromosphere'' models have larger critical radii that
are in the super-Alfv\'{e}nic ($u > V_{\rm A}$) part of the wind.
Also plotted in Figure 8{\em{a}} is the so-called wind luminosity
$L_{\rm wind}$, which we estimate from the sum of the energy required
to lift the wind out of the gravitational potential and the
remaining kinetic energy of the flow, i.e.,
\begin{equation}
  L_{\rm wind} \, = \, \dot{M}_{\rm wind} \left(
  \frac{GM_{\ast}}{R_{\ast}} + \frac{u^2}{2} \right)
  \label{eq:Lwind}
\end{equation}
(e.g., Clayton 1966).
This expression ignores thermal energy, ionization energy,
magnetic energy, and waves, which are expected to be small
contributors at and above the critical point.
Curves for the ratio $L_{\rm wind}/L_{\ast}$ are shown both for
the wind at its largest height (i.e., using $u_{\infty}$ in
eq.~[\ref{eq:Lwind}]) and at the critical point (using
$u_{\rm crit}$).
The latter is helpful to compare with analytic estimates for
younger ages described below---for which nothing above the
critical point was computed.

Figure 8{\em{b}} shows the terms that dominate the energy
conservation equation (at the critical point) as a function of age.
The areas plotted here were computed by normalizing the absolute
values of the individual terms in equation (\ref{eq:dEdt}) by the
maximum absolute value for each model and then ``stacking'' them so
that together they fill the region between 0 and 1.
The older coronal models have a three-part balance between
Alfv\'{e}n wave heating, heat conduction, and the upward advection
of enthalpy due to the terms on the left-hand side of
equation (\ref{eq:dEdt}).
For the younger models, radiative losses become important because
of the higher densities at the critical point, and heat conduction
disappears because the temperatures are so low.

\begin{figure}
\epsscale{1.05}
\plotone{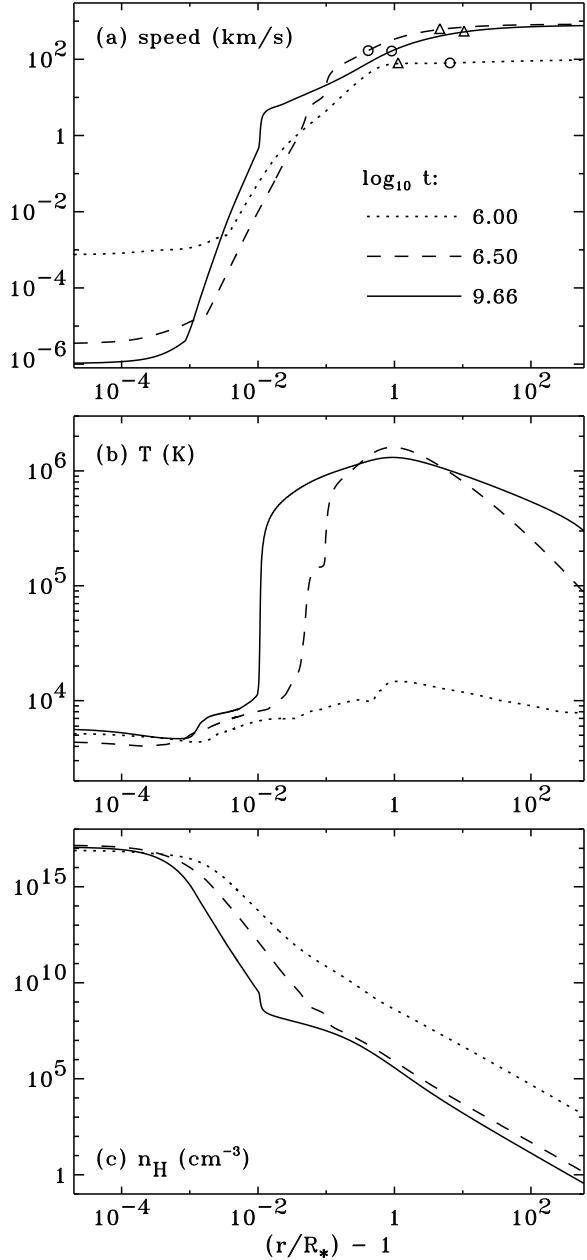}
\caption{Radial dependence of wind parameters for three selected
ZEPHYR models:
({\em{a}}) wind outflow velocity, ({\em{b}}) temperature, and
({\em{c}}) total hydrogen number density.
In all panels, models are shown for ages $\log_{10} t = 6.0$
({\em dotted lines}), 6.5 ({\em dashed lines}), and 9.66
({\em solid lines}).
Wave-modified critical radii ({\em circles}) and
Alfv\'{e}n radii ({\em triangles}) are also shown.}
\end{figure}

Figure 9 displays the radial dependence of wind speed, temperature,
and density for three selected ages: (1) the present-day polar
outflow of the Sun, (2) a younger, but still low-$\dot{M}_{\rm wind}$
coronal wind, and (3) an even younger model that has made the
transition to a higher mass loss rate and an extended chromosphere.
The plotted hydrogen number density $n_{\rm H}$ is that of all
hydrogen nuclei (both neutral and ionized).
Note that the model with the lowest temperature would be expected
to have a small density scale height (eq.~[\ref{eq:Hast}]) and
thus a rapid radial decline in $n_{\rm H}$.
However, Figure 9{\em{c}} shows the opposite to be the case:
the coolest model has the largest {\em effective} density scale height
of the three because of a much larger contribution from wave pressure.

There are two additional points of comparison with other work that
should be made in the light of the main results shown in Figures 7--9.
\begin{enumerate}
\item
There is a relatively flat age dependence of the predicted mass loss
rate for the post-T~Tauri phase (i.e., zero-age main sequence [ZAMS]
stars with $t \gtrsim 50$ Myr).
This stands in contrast to the observationally inferred power-law
decrease for these stages, which is approximately
$\dot{M}_{\rm wind} \propto t^{-2}$ (e.g., Wood et al.\  2002, 2005).
Note, however, that the internal (convection-related) source of
MHD wave energy at the photosphere was assumed in the models to be
set at the present-day solar level and not to vary with age.
As described in {\S}~6, the inclusion of {\em rotation-dependent}
convection may give rise to a stronger variation in the
turbulent fluctuation amplitude with age, and thus also a more
pronounced age dependence of the rates of coronal heating and
mass loss.
\item
Suzuki (2007) modeled the extended atmospheres and winds of 
low-gravity cool stars, and found that the onset of thermal
instability gave rise to large amounts of time variability.
These models often showed dense and cool shells that coexist with
spontaneously produced hot and tenuous ``bubbles.''
Suzuki (2007) noted that this dynamical instability began to occur
once the escape velocity $V_{\rm esc}$ (measured in the wind
acceleration region) dropped to the point of being of the same
order of magnitude as the sound speed corresponding to the
thermal instability temperature of $\sim 10^5$ K.
It is unclear, though, whether this variability is triggered only
for relatively moderate rates of turbulent energy input, like the
amplitudes derived by Suzuki (2007) from non-rotating convection
models.
The larger amplitudes used in the present models of accretion-driven
waves may be sufficient to drive the winds much more robustly
{\em past} the thermal instability and into time-steady
extended chromospheres (see also Killie 2002).
\end{enumerate}

\subsection{Disappearance of Time-Steady Solutions}

The ZEPHYR code could not find time-independent wind solutions for
ages younger than about 0.45 Myr (i.e., for accretion rates larger
than about $7 \times 10^{-8}$ $M_{\odot}$ yr$^{-1}$).
This was found {\em not} to be a numerical effect.
Instead, it is an outcome of the requirement that time-steady winds
must have a sufficient amount of outward acceleration (either due
to gas pressure or some other external forcing, like waves) to
drive material out of the star's gravitational potential well
to a finite coasting speed at infinity.
This was realized for {\em polytropic} gas-pressure-driven winds
very early on (Parker 1963; Holzer \& Axford 1970),
such that if the temperature decreases too rapidly with increasing
distance (i.e., with decreasing density) the wind would become
``stalled'' and not have a time-steady solution.

To illustrate how this effect occurs for the youngest, wave-driven
models, let us write and analyze an approximate equation of momentum
conservation.
For simplicity, the wind temperature $T$ is assumed to be constant,
and the radii of interest are far enough from the star such that
the flux-tube expansion can be assumed to be spherical, with
$A \propto r^{2}$.
Also, for these models the acoustic wave pressure can be ignored
(since compressive waves damp out rather low in the atmosphere), and
the radial behavior of the Alfv\'{e}n wave amplitude can be modeled
roughly in the dissipationless limit.
Thus, one can follow Jacques (1977) and write the
momentum equation as a modified critical point equation
\begin{equation}
  \left( u - \frac{u_{\rm crit}^2}{u} \right) \frac{du}{dr} \, = \,
  - \frac{GM_{\ast}}{r^2} + \frac{2 u_{\rm crit}^{2}}{r} \,\, .
  \label{eq:ucold}
\end{equation}
At the wave-modified critical point ($r_{\rm crit}$), the wind speed
$u$ equals the critical speed $u_{\rm crit}$, which is defined as
\begin{equation}
  u_{\rm crit}^{2} \, = \, a^{2} + \frac{v_{\perp}^2}{4} \left(
  \frac{1 + 3 M_{\rm A}}{1 + M_{\rm A}} \right)
  \label{eq:ucrit}
\end{equation}
where the squared isothermal sound speed is
$a^{2} = (1+x) k_{\rm B}T/m_{\rm H}$,
and the bulk-flow Alfv\'{e}n Mach number $M_{\rm A} = u/V_{\rm A}$.
A more general version of equation (\ref{eq:ucold}) that also contains
damping and acoustic wave pressure is given in, e.g., {\S}~6 of
Cranmer et al.\  (2007).

The above expressions show how the outward pressure, which balances
gravity at the critical point, can be dominated either by $a^2$
(gas pressure) or by a term proportional to $v_{\perp}^{2}$
(wave pressure).
For gas pressure that can be described as a polytrope (i.e.,
$a^{2} \propto \rho^{\gamma-1}$), the polytropic index $\gamma$ at
the critical point must be smaller than 1.5 in order for there to
be a time-steady acceleration from the critical point to infinity
(Parker 1963; see also Velli 2001; Owocki 2004).
Larger polytropic indices $\gamma \geq 1.5$ imply that $a^2$ drops
too rapidly with increasing distance to provide sufficient acceleration
for a parcel accelerated through the critical point to escape
to infinity.

For winds dominated by wave pressure, it is possible to use the
equation of wave action conservation (eq.~[\ref{eq:wact}])
to examine the density dependence of the wave amplitude in a similar
way as above (e.g., Jacques 1977; Heinemann \& Olbert 1980;
Cranmer \& van Ballegooijen 2005).
The exponent $\xi$ in the scaling relation
$v_{\perp} \propto \rho^{\xi}$ is known to be a slowly varying
function of distance.
Close to the star, where $u \ll V_{\rm A}$, the exponent
$\xi \approx -0.25$.
In the vicinity of the Alfv\'{e}n point ($u \approx V_{\rm A}$),
$\xi$ increases to zero and $v_{\perp}(r)$ has a local maximum.
Far from the star, where $u \gg V_{\rm A}$, the exponent $\xi$
grows to an asymptotically large value of $+0.25$.
The dimensionless quantity in parentheses in equation (\ref{eq:ucrit})
varies only between 1 and 3 over the full range of distances, and can
be assumed to be roughly constant compared to the density
dependence of $v_{\perp}$.

Thus, by comparing the polytropic, gas-pressure dominated expression
for $u_{\rm crit}^2$ to the wave-pressure dominated version, it
becomes possible to write an {\em effective polytropic index} for
the latter case as $\gamma_{\rm eff} = 2\xi + 1$.
The unstable region of $\gamma \geq 1.5$ corresponds to $\xi \geq 0.25$,
and this value is reached at the critical point only when
$r_{\rm crit}$ is well into the super-Alfv\'{e}nic part of the wind.
Indeed, this corresponds to the youngest models shown in Figures 7--9.
As $\dot{M}_{\rm wind}$ increases (with decreasing age), the density
in the wind increases, and this leads to a sharp decline in the value
of $V_{\rm A}$ at the critical point.
(Just over the span of ages going from $\log_{10} t = 6.25$ to 6.0,
$V_{\rm A}$ at the critical point drops by a factor of 150.)
Figure 8{\em{b}} shows $\xi$ versus age for the ZEPHYR models, where
this exponent was first computed for all heights via
\begin{equation}
  \xi \, = \, \frac{\partial (\ln v_{\perp}) / \partial r}
  {\partial (\ln \rho) / \partial r}
\end{equation}
and the value shown is that interpolated to the location of the
wave-modified critical point.
It is clear that $\xi$ approaches the unstable limiting value of 0.25
just at the point where the time-steady solutions disappear.

What happens to a stellar outflow when ``too much'' mass is driven
up past the critical point to maintain a time-steady wind?
An isolated parcel of gas with an outflow speed $u = u_{\rm crit}$
at the critical point would be decelerated to stagnation at some
height above the critical point, and it would want to fall back
down towards the star.
In reality, this parcel would collide with other parcels that
are still accelerating, and a stochastic collection of shocked
clumps is likely to result.
Interactions between these parcels may result in an extra degree
of collisional heating that could act as an extended source of
gas pressure to help maintain a mean net outward flow.
Situations similar to this have been suggested to occur in the
outflows of both pulsating cool stars (e.g., Bowen 1988;
Willson 2000; Struck et al.\  2004) and luminous blue variables
(Owocki \& van Marle 2008).
The models presented here suggest that the most massive stellar
winds ($\dot{M}_{\rm wind} \gtrsim 10^{-9}$ $M_{\odot}$ yr$^{-1}$)
of young T Tauri stars may exist in a similar kind of superposition
of outflowing and inflowing shells.

\subsection{Analytic Estimate of Wave-Driven Mass Loss}

For the youngest T Tauri models with (seemingly) no steady state,
it is possible to use an analytic technique to estimate how much
mass gets accelerated up to the wave-modified critical point.
As described above, it is not certain whether all of this mass can
be accelerated to infinity.
However, the ability to determine a mass flux that applies to the
finite region between the stellar surface and the critical radius
may be sufficient to predict many observed mass loss diagnostics.
Figure 8{\em{a}} showed that $r_{\rm crit} \gg R_{\ast}$
for these young ages, and thus the ``subcritical volume'' is
relatively large.

The modified critical point equation given above for the limiting
case of dissipationless Alfv\'{e}n waves has been studied for
several decades (e.g., Jacques 1977; Hartmann \& MacGregor 1980;
DeCampli 1981; Holzer et al.\  1983; Wang \& Sheeley 1991).
Equation (\ref{eq:ucold}) can be solved for the critical point radius
by setting the right-hand side to zero, to obtain
\begin{equation}
  r_{\rm crit} \, = \, \frac{GM_{\ast}}{2 u_{\rm crit}^2}
  \,\, , \,\,\,\,\,
  u_{\rm crit}^{2} \, = \, a^{2} + \frac{3 v_{\perp {\rm crit}}^2}{4}
\end{equation}
where it is assumed that $u \gg V_{\rm A}$ (or $M_{\rm A} \gg 1$) at
the critical point, which applies in the youngest T Tauri models.
In order to compute a value for the critical point radius, however,
we would have to know the value of $v_{\perp {\rm crit}}$, the
Alfv\'{e}n wave amplitude at the critical point.
No straightforward (a~priori) method was found to predict
$v_{\perp {\rm crit}}$ from the other known quantities derived
in {\S}~3.
Thus, we relied on an empirical fitting relation that was produced
from the 6 youngest ZEPHYR models (with $0.45 \leq t \leq 1.4$ Myr),
\begin{equation}
  \frac{v_{\perp {\rm crit}}}{18.9 \,\, \mbox{km} \,\, \mbox{s}^{-1}}
  \, = \, \left(
  \frac{v_{\perp \ast}}{10 \,\, \mbox{km} \,\, \mbox{s}^{-1}}
  \right)^{-2.37}
\end{equation}
where $v_{\perp \ast}$ is the Alfv\'{e}n wave amplitude at the
photosphere.
The fit is good to 5\% accuracy for the 6 youngest numerical models,
but it is not known whether this level of accuracy persists when
it is extrapolated to ages younger than $\sim$0.45 Myr.

The assumption that the Alfv\'{e}n waves are dissipationless allows
the conservation of wave action to be used, i.e.,
\begin{equation}
  \frac{\rho v_{\perp}^{2} (u + V_{\rm A})^{2} A}{V_{\rm A}} \, = \,
  \mbox{constant}  \,\, .
  \label{eq:wact}
\end{equation}
This can be simplified by noting that, for the youngest T Tauri
models, the photosphere always exhibits $u \ll V_{\rm A}$, and the
wave-modified critical point always exhibits $u \gg V_{\rm A}$.
Thus, with these assumptions, equation (\ref{eq:wact}) can be
applied at these two heights to obtain
\begin{equation}
  \rho_{\ast} v_{\perp \ast}^{2} V_{{\rm A} \ast} A_{\ast}
  \, \approx \,
  \frac{\rho_{\rm crit} v_{\perp {\rm crit}}^{2}
  u_{\rm crit}^{2} A_{\rm crit}}{V_{\rm A \, crit}}  \,\, .
\end{equation}
The main unknown is the density $\rho_{\rm crit}$ at the critical
point, which can be solved in terms of the other known quantities,
\begin{equation}
  \left( \frac{\rho_{\rm crit}}{\rho_{\ast}} \right)^{3/2}
  \, \approx \,
  \frac{B_{\rm crit}^2}{4\pi \rho_{\ast} u_{\rm crit}^2}
  \left( \frac{v_{\perp \ast}}{v_{\perp {\rm crit}}} \right)^{2}
\end{equation}
and thus the mass flux is determined uniquely at the critical point
(Holzer et al.\  1983).
The mass loss rate is
$\dot{M}_{\rm wind} = \rho_{\rm crit} u_{\rm crit} A_{\rm crit}$,
where the critical flux-tube area $A_{\rm crit}$ is normalized such
that as $r \rightarrow \infty$, $A \rightarrow 4\pi r^{2}$.

Although the above provides a straightforward algorithm for estimating
$\dot{M}_{\rm wind}$ for a wave-driven wind, there is one other
unspecified quantity: the isothermal sound speed $a$ at the
critical point.
A completely ``cold'' wave-driven model would set $a = 0$, but
it was found that the extended chromospheres in the youngest ZEPHYR
models do contribute some gas pressure to the overall acceleration.
The present set of approximate models began with an initial guess
for the critical point temperature ($T_{\rm crit} \approx 10^4$ K)
and then iterated to find a more consistent value.
The iteration process involved alternating between solving for
$\dot{M}_{\rm wind}$ at the critical point (as described above)
and recomputing $T_{\rm crit}$ by assuming a balance between
Alfv\'{e}n-wave turbulent heating and radiative cooling at
the critical point.
The turbulent heating rate $Q_{\rm A}$ was computed as in equation
(\ref{eq:Qdmit}), and the cooling rate $Q_{\rm rad}$ was assumed to
remain proportional to its optically thin limit $\rho^{2} \Lambda(T)$
(i.e., ignoring the $\tau_{\rm hk}$ correction factor described in
{\S}~4.1).
The temperature at which this balance occurred was found by
inverting the tabulated radiative loss function $\Lambda(T)$,
shown in Figure 1 of Cranmer et al.\  (2007).
For each age---between 13.5 kyr and 0.45 Myr---this process was run
for 20 iterations, but in all cases it converged rapidly within the
first 5 iterations.

The results of this iterative estimation of $\dot{M}_{\rm wind}$,
$T_{\rm crit}$, and $u_{\rm crit}$ are shown by the dotted curves
in Figures 7 and 8.
The maximum mass loss rate of $2.4 \times 10^{-8}$
$M_{\odot}$ yr$^{-1}$ occurs at an age of about 40 kyr.
At the youngest ages, the mass loss rate declines because the
accretion-driven wave power also declines (see Fig.~5).
The most massive wind corresponds to the minimum value of the
temperature at the critical point, which is indistinguishable
from the stellar $T_{\rm eff}$ at that age.
At the oldest ages for which these solutions were attempted
($t \gtrsim 1.5$ Myr), the various approximations made above no
longer hold (e.g., the critical point is no longer
super-Alfv\'{e}nic) and the agreement between the analytic
estimates and the ZEPHYR models breaks down.

\subsection{Varying the Accretion Rate}

As shown in Figure 3 above, the measured values for $\dot{M}_{\rm acc}$
exhibit a wide spread around the mean relation that was used to
derive the properties of accretion-driven waves.
In order to explore how variability in the accretion rate affects
the resulting stellar wind, a set of additional ZEPHYR models was
constructed with a factor of two decrease and increase in
$\dot{M}_{\rm acc}$ in comparison to equation (\ref{eq:Maccfit}).
These models were constructed mainly for ages around the thermal
bifurcation at $\log_{10} t \approx 6.25$.
It was found that the older ``hot corona'' models were insensitive
to variations in the accretion rate, even at ages where the
accretion-driven component to $v_{\perp\ast}$ exceeded
the internal convective component.

\begin{figure}
\epsscale{1.05}
\plotone{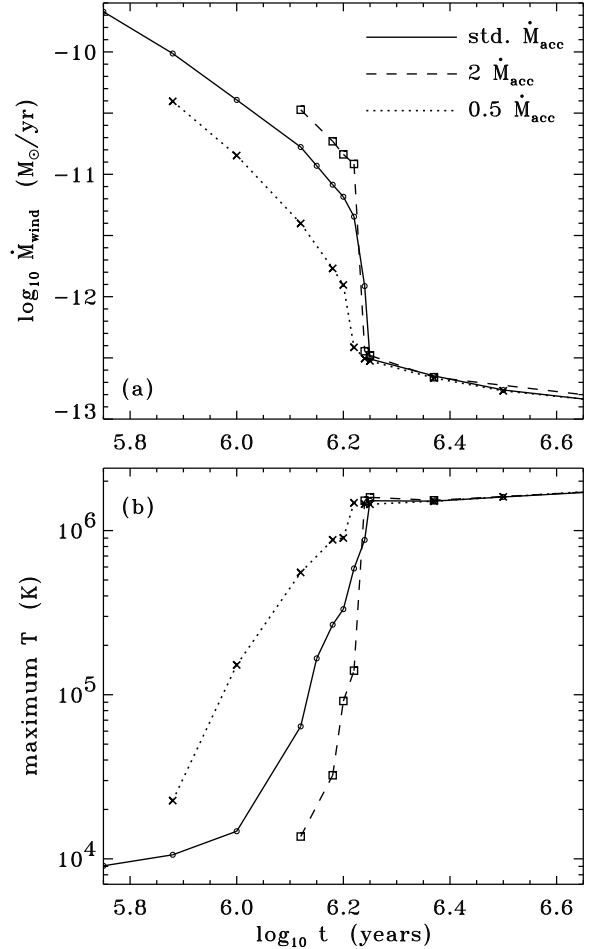}
\caption{Age-dependent mass loss rates ({\em{a}}) and peak wind
temperatures ({\em{b}}) for ZEPHYR models produced with the
standard accretion rate from eq.~(\ref{eq:Maccfit})
({\em solid lines}), and accretion rates that are double
({\em dashed lines}) and half ({\em dotted lines}) the
standard values.
Individual ZEPHYR models are denoted by symbols.}
\end{figure}

Figure 10 compares the mass loss rate and maximum wind temperature
for the three sets of ZEPHYR models: the original standard
models, ones with half of the accretion rate given by
equation (\ref{eq:Maccfit}), and ones with double that rate.
The specific age at which the thermal bifurcation occurs changes by
only a small amount over the range of modeled accretion rates.
(Indeed, the standard model and the double-$\dot{M}_{\rm acc}$
model undergo thermal bifurcation at nearly the exact same age.)
The youngest and coolest models show a greater degree of responsiveness
to the varying accretion rate than do the older hot models.
The younger models with higher accretion rates have larger
photospheric MHD wave amplitudes, and thus they give rise to larger
mass loss rates and cooler extended chromospheres.

\begin{figure}
\epsscale{1.05}
\plotone{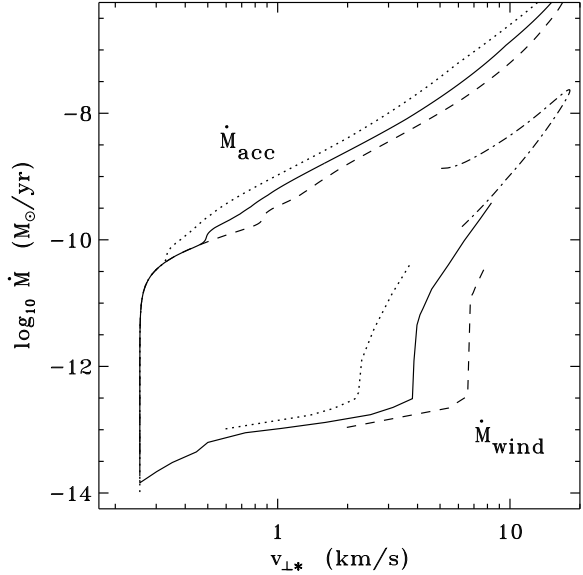}
\caption{Mass accretion and mass loss rates for the 3 sets of
ZEPHYR models shown in Figure 10 (with the same line styles) shown
versus the photospheric Alfv\'{e}n wave amplitude $v_{\perp\ast}$.
Mass loss rates for the analytic/cold models of {\S}~5.3 are also
shown for the standard accretion rate case ({\em dot-dashed line}).}
\end{figure}

Figure 11 shows how both $\dot{M}_{\rm acc}$ and
$\dot{M}_{\rm wind}$ vary as a function of the photospheric
Alfv\'{e}n wave amplitude $v_{\perp\ast}$.
This latter quantity is a key intermediary between the mid-latitude
accretion and the polar mass loss, and thus it is instructive to see
how both mass flux quantities scale with its evolving magnitude.
The three sets of ZEPHYR models from Figure 10 are also shown in
Figure 11, and it is noteworthy that there is {\em not} a simple
one-to-one correspondence between $\dot{M}_{\rm acc}$ and
$v_{\perp\ast}$.
The small spread arises because the fundamental stellar parameters
(e.g., $R_{\ast}$ and $\rho_{\ast}$) are different at the three
ages that correspond to the same value of $\dot{M}_{\rm acc}$ in
the three models.
Thus, it is not surprising that $\dot{M}_{\rm wind}$ is not a
``universal'' function of the wave amplitude either.

\begin{figure}
\epsscale{1.05}
\plotone{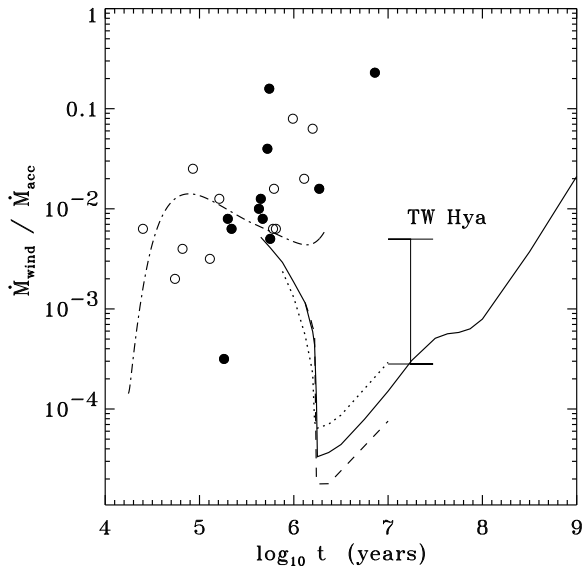}
\caption{Ratio of mass loss rate to mass accretion rate for
measured T Tauri stars (symbols as in Fig.~3) and for the ZEPHYR
and analytic models (line styles as in Figs.~10 and 11).}
\end{figure}

Figure 12 plots the key ``wind efficiency'' ratio
$\dot{M}_{\rm wind}/\dot{M}_{\rm acc}$ as a function of age for
the three sets of ZEPHYR models and the analytic estimates derived
in {\S}~5.3.
Note that, in contrast to Figure 10, when plotted as a ratio,
the models on the young/cool side of the thermal bifurcation all
seem to collapse onto a single curve, while the older/hotter models
separate (based on differing accretion rates in the denominator).

For both the youngest ages ($t \lesssim 0.5$ Myr) and for the specific
case of TW Hya, the model values shown in Figure 12 seem to agree
reasonably well with the observationally determined ratios that
are also shown in Figure 3{\em{b}}.
The models having ages between 0.5 and 10 Myr clearly fall well below
the measured mass loss rates.
However, even the limited agreement with the data is somewhat
surprising, since these measured values come from the [\ion{O}{1}]
$\lambda$6300 forbidden line diagnostic that is widely believed
to sample the much larger-scale disk wind (Hartigan et al.\  1995).
It is thus possible that stellar winds may contribute to
observational signatures that previously have been assumed to
probe only the (disk-related) bipolar jets.

\subsection{X-Ray Emission}

Many aspects of the dynamics and energetics of young stars and their
environments are revealed by high-energy measurements such as X-ray
emission (e.g., Feigelson \& Montmerle 1999).
It is thus worthwhile to determine the level of X-ray flux that the
modeled polar winds are expected to generate.
This has been done in an approximate way to produce order-of-magnitude
estimates, and should be followed up by more exact calculations in
future work.

The optically thin radiative loss rate $Q_{\rm rad}$ described in
{\S}~4.1 was used as a starting point to ``count up'' the total
number of photons generated by each radial grid zone in the ZEPHYR
models.
(Finite optical depth effects in $Q_{\rm rad}$ were ignored here
because they contribute mainly to temperatures too low to affect
X-ray fluxes.)
This rate depends on the plasma temperature $T$, the density $\rho$,
and the hydrogen ionization fraction $x$.
The total radiative loss rate is multiplied by a fraction $F$ that
gives only those photons that would be observable as X-rays.
This fraction is estimated for each radial grid zone as
\begin{equation}
  F \, = \, \frac{\int d\lambda \,\, B_{\lambda}(T) \, S(\lambda)}
  {\int d\lambda \,\, B_{\lambda}(T)}
\end{equation}
where $B_{\lambda}(T)$ is the Planck blackbody function and
$S(\lambda)$ is an X-ray sensitivity function, for which we use that
of the {\em ROSAT} Position Sensitive Proportional Counter (PSPC)
instrument, as specified by Judge et al.\  (2003).
The function $S(\lambda)$ is nonzero between about 0.1 and 2.4 keV,
with a minimum around 0.3 keV that separates the hard and soft bands.
The integration in the denominator is taken over all wavelengths.
The use of the blackbody function, rather than a true optically
thin emissivity, was validated by comparison with wavelength-limited
X-ray radiative loss rates given by Raymond et al.\  (1976).
Fractions of emissivity (relative to the total loss rate) in specific
X-ray wavebands were computed for $T = 0.5,$ 1, 2, and 5 MK
and compared with the plots of Raymond et al.\  (1976).
The agreement between the models and the published curves was always
better than a factor of two.\footnote{%
Although this is obviously not accurate enough for quantitative
comparisons with specific observations, it allows the correct order of
magnitude of the X-ray emission to be estimated; see Figure 13.
Also, the factor-of-two validation should be taken in the context
of the factor of $\sim$1000 variation in these fractions over the
modeled coronal temperatures.}

\begin{figure}
\epsscale{1.05}
\plotone{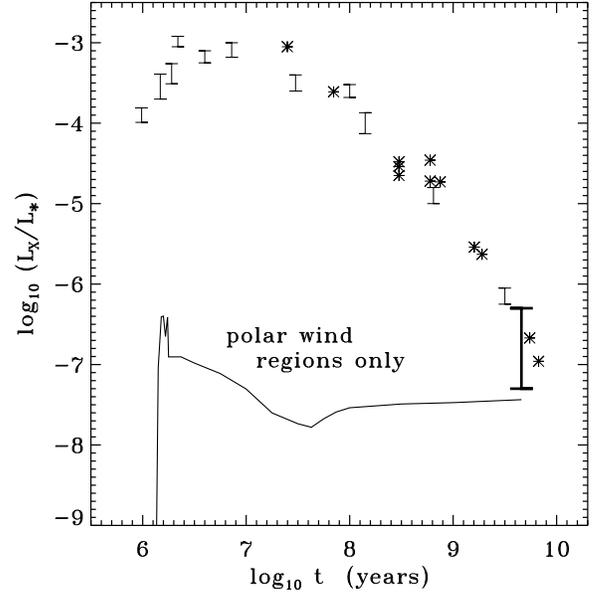}
\caption{Ratio of X-ray luminosity to total bolometric luminosity
for the standard run of ZEPHYR models ({\em solid line}), and for
observations of solar-type stars ({\em asterisks}), nearby clusters
({\em thin error bars}), and the Sun ({\em thick error bar}); see
text for details.}
\end{figure}

Figure 13 shows the simulated ratio of X-ray luminosity $L_{\rm X}$
to the bolometric luminosity $L_{\ast}$ for the modeled wind regions.
For each ZEPHYR model, the radiative losses were integrated over
an assumed spherical volume for the stellar wind (i.e., the same
assumption used to compute $\dot{M}_{\rm wind}$; see {\S}~5.1) to
produce $L_{\rm X}$.  No absorption of X-rays was applied.
Figure 13 also shows a collection of observed X-ray luminosity
ratios for individual solar-type stars
(from G\"{u}del et al.\  1998; Garc\'{\i}a-Alvarez et al.\  2005)
and clusters of various ages (Flaccomio et al.\  2003b;
Preibisch \& Feigelson 2005; Jeffries et al.\  2006).
For the latter, the error bars indicate the $\pm 1 \sigma$ spread
about the mean values reported in these papers.
The range of values for the present-day Sun is taken from
Judge et al.\  (2003).

It is clear from Figure 13 that the modeled polar wind regions do
not produce anywhere near enough X-rays to explain the observations
of young stars.
For the present-day Sun, the computed value slightly underestimates
the lower limit of the observed range of X-ray luminosities.
This latter prediction does make some sense, since the X-ray emission
is essentially computed for the dark ``polar coronal holes.''
On the Sun, these never occupy more than about 20\% of the surface.
Most notably, though, there is no strong power-law decrease in
$L_{\rm X}/L_{\ast}$ for young ZAMS stars---just as there is no
power-law decrease in $\dot{M}_{\rm wind}$---because there has been
no attempt to model the rotation-age-activity relationship
(Skumanich 1972; Noyes et al.\  1984).
The ZEPHYR models neglect the closed-field coronae of these stars
(both inside and outside the polar-cap regions) that are likely to
dominate the X-ray emission, like they do for the Sun
(e.g., Schwadron et al.\  2006).

It is somewhat interesting that the ZEPHYR models undergo the
thermal bifurcation to extended chromospheres (and thus disappear
from Figure 13 because of the lack of X-ray emitting temperatures)
at about the same age where there seems to exist a slight decline
in the X-ray emission of young T Tauri stars.
Although there is some evidence for such a deficit of X-rays in
classical T Tauri stars, with respect to the older class of
weak-lined T Tauri stars (e.g., Flaccomio et al.\  2003a,b;
Telleschi et al.\  2007), the existence of a distinct peak in X-ray
emission at intermediate ages has not been proven rigorously.
If the polar wind regions indeed produce negligible X-ray emission
compared to the closed-field regions, we note that the polar cap area
$\delta_{\rm pol}$ grows largest for the youngest T Tauri stars
(see Fig.~4).
Thus, the cooler wind material may occupy a significantly larger
volume than the hotter (closed-loop) coronal plasma at the youngest
ages.  This may be partly responsible for the observed X-ray trends.

\section{Discussion and Conclusions}

The primary aim of this paper has been to show how accretion-driven
waves on the surfaces of T Tauri stars may help contribute to the
strong rates of atmospheric heating and large mass loss rates
inferred for these stars.
The ZEPHYR code, which was originally developed to model the solar
corona and solar wind (Cranmer et al.\  2007), has been applied to
the T Tauri stellar wind problem.
A key aspect of the models presented above is that the only true free
parameters are: (1) the properties of MHD waves injected at the
photospheric lower boundary, and (2) the background magnetic geometry.
Everything else (e.g., the radial dependence of the rates of
chromospheric and coronal heating, the temperature structure of the
atmosphere, and the wind speeds and mass fluxes) emerges naturally
from the modeling process.

For solar-mass T Tauri stars, time-steady stellar winds were found to
be supportable for all ages older than about 0.45 Myr, with accretion
rates less than $7 \times 10^{-8}$ $M_{\odot}$ yr$^{-1}$ driving
mass loss rates less than $4 \times 10^{-10}$ $M_{\odot}$ yr$^{-1}$.
Still younger T Tauri stars (i.e., ages between about 13 kyr and
0.45 Myr) may exhibit time-variable winds with mass loss rates
extending up several more orders of magnitude to
$\sim 2 \times 10^{-8}$ $M_{\odot}$ yr$^{-1}$.
The transition between time-steady and variable winds occurs when the
critical point of the flow migrates far enough past the Alfv\'{e}n
point (at which the wind speed equals the Alfv\'{e}n speed) such
that the Alfv\'{e}n wave amplitude begins to decline rapidly with
increasing radius.
When this happens, the outward wave-pressure acceleration is quickly
``choked off;'' i.e., parcels of gas that make it past the critical
point cannot be accelerated to infinity, and stochastic collisions
between upflowing and downflowing parcels must begin to occur.

The maximum wind efficiency ratio
$\dot{M}_{\rm wind}/\dot{M}_{\rm acc}$ for the T Tauri models
computed here was approximately 1.4\%, computed for ages of order
0.1 Myr.
This is somewhat smaller than the values of order 10\% required by
Matt \& Pudritz (2005, 2007, 2008) to remove enough angular momentum
from the young solar system to match present-day conditions.
It is also well below the observational ratios derived by, e.g.,
Hartigan et al.\  (1995, 2004), which can reach up to 20\% for
T Tauri stars between 1 and 10 Myr old (see Fig.~12).
These higher ratios, though, may be the product of both stellar
winds and disk winds (possibly even dominated by the disk wind
component).
Additionally, it is possible that future observational analysis
will result in these empirical ratios being revised {\em upward}
with more accurate (lower) values of $\dot{M}_{\rm acc}$
(S.\  Edwards 2008, private communication).

The accretion-driven solutions for $\dot{M}_{\rm wind}$ depend
crucially on the properties of the waves in the polar regions.
It is important to note that the calculation of MHD
wave properties was based on several assumptions that should
be examined in more detail.
The relatively low MHD wave efficiency used in equation (\ref{eq:EA})
is an approximation based on the limiting case of waves being far
from the impact site.
A more realistic model would have to contain additional information
about both the nonlinearities of the waves themselves and the
vertical atmospheric structure through which the waves propagate.
It seems likely that a better treatment of the wave generation would
lead to larger wave energies at the poles.
On the other hand, the assumption that the waves do not damp
significantly between their generation point and the polar wind
regions may have led to an assumed wave energy that is too high.
It is unclear how strong the waves will be in a model that takes
account of all of the above effects.
In any case, the ZEPHYR results presented here are the first
self-consistent models of T Tauri stellar winds that produce
wind efficiency ratios that even get into the right ``ballpark''
of the angular momentum requirements.

Additional improvements in the models are needed to make further progress.
For example, the effects of stellar rotation should be included, both in
the explicit wind dynamics (e.g., ``magneto-centrifugal driving,'' as
recently applied by Holzwarth \& Jardine 2007) and in the modified
subsurface convective activity that is likely to affect photospheric
amplitudes of waves and turbulence.
Young and rapidly rotating stars are likely to have qualitatively
different convective motions than are evident in standard
(nonrotating, mixing-length) models.
It is uncertain, though, whether rapid rotation gives rise to larger
(K\"{a}pyl\"{a} et al.\  2007; Brown et al. 2007; Ballot et al. 2007)
or smaller (Chan 2007) convection eddy velocities at the latitudes
of interest for T Tauri stellar winds.
In any case, rapid rotation can also increase the buoyancy of
subsurface magnetic flux elements, leading to a higher rate of flux
emergence (Holzwarth 2007).
Also, the lower gravities of T Tauri stars may give rise to a larger
fraction of the convective velocity reaching the surface as wave
energy (e.g., Renzini et al.\  1977), or the convection may even
penetrate directly into the photosphere (Montalb\'{a}n et al.\ 2004).

Future work must involve not only increased physical realism for the
models, but also quantitative comparisons with observations.
The methodology outlined in this paper should be applied to a set
of real stars, rather than to the idealized evolutionary sequence
of representative parameters.
Measured stellar masses, radii, and $T_{\rm eff}$ values, as well as
accretion rates, magnetic field strengths, and hot spot filling factors
($\delta$), should be used as constraints on individual calculations
for the stellar wind properties.
It should also be possible to use measured three-dimensional magnetic
fields (e.g., Donati et al.\  2007; Jardine et al.\  2008) to more
accurately map out the patterns of accretion stream footpoints, wave
fluxes, and the flux tubes in which stellar winds are accelerated.

This work helps to accomplish the larger goal of understanding the
physics responsible for low-mass stellar outflows and the feedback
between accretion disks, winds, and stellar magnetic fields.
In addition, there are links to more interdisciplinary studies of
how stars affect objects in young solar systems.
For example, the coronal activity and wind of the young Sun is likely to
have created many of the observed abundance and radionuclide patterns in
early meteorites (Lee et al.\  1998) and possibly affected the Earth's
atmospheric chemistry to the point of influencing the development of
life (see, e.g., Ribas et al.\  2005; G\"{u}del 2007;
Cuntz et al.\  2008).
The identification of key physical processes in young stellar winds
is important not only for understanding stellar and planetary evolution,
but also for being able to model and predict the present-day impacts of
solar variability and ``space weather'' (e.g.,
Feynman \& Gabriel 2000; Cole 2003).

\acknowledgments

I gratefully acknowledge Andrea Dupree, Adriaan van Ballegooijen,
Nancy Brickhouse, and Eugene Avrett for many valuable discussions.
This work was supported by the National Aeronautics and
Space Administration (NASA) under grant {NNG\-04\-GE77G}
to the Smithsonian Astrophysical Observatory.


\begin{thebibliography}{}

\bibitem[]{Air0} Airapetian, V. S., Ofman, L., Robinson, R. D.,
  Carpenter, K., \& Davila, J. 2000, \apj, 528, 965

\bibitem[]{An86} Antiochos, S. K., Haisch, B. M., \& Stern, R. A.
  1986, \apjl, 307, L55

\bibitem[]{AM89} Appenzeller, I., \& Mundt, R. 1989, \aapr, 1, 291

\bibitem[]{Ar07} Argiroffi, C., Maggio, A., \& Peres, G. 2007,
  \aap, 465, L5

\bibitem[]{As06} Aschwanden, M. J. 2006, Physics of the Solar
  Corona: An Introduction with Problems and Solutions, 2nd ed.
  (Berlin: Springer)

\bibitem[]{AL08} Avrett, E. H., \& Loeser, R. 2008, \apjs, 175, 229

\bibitem[]{Bk07} Balasubramaniam, K. S., Pevtsov, A. A., \&
  Neidig, D. F. 2007, \apj, 658, 1372

\bibitem[]{Ba07} Ballot, J., Brun, A. S., \& Turck-Chi\`{e}ze, S.
  2007, \apj, 669, 1190

\bibitem[]{Bt02} Batalha, C., Batalha, N. M., Alencar, S. H. P.,
  Lopes, D. F., \& Duarte, E. S. 2002, \apj, 580, 343

\bibitem[]{Bc71} Belcher, J. W. 1971, \apj, 168, 509

\bibitem[]{Bt89} Bertout, C. 1989, \araa, 27, 351

\bibitem[]{Bs08} Bessolaz, N., Zanni, C., Ferreira, J., Keppens, R.,
  \& Bouvier, J. 2008, \aap, 478, 155

\bibitem[]{Bu04} Bouvier, J., Dougados, C., \& Alencar, S. H. P.
  2004, \apss, 292, 659

\bibitem[]{Bu97} Bouvier, J., et al. 1997, \aap, 318, 495

\bibitem[]{Bu03} Bouvier, J., et al. 2003, \aap, 409, 169

\bibitem[]{Bu07} Bouvier, J., et al. 2007, \aap, 463, 1017

\bibitem[]{Bw88} Bowen, G. H. 1988, \apj, 329, 299

\bibitem[]{BG68} Bretherton, F. P., \& Garrett, C. J. R. 1968,
  Proc.\  Roy.\  Soc.\  A, 302, 529

\bibitem[]{Br07} Brown, B. P., Browning, M., Brun, A. S., Miesch,
  M. S., \& Toomre, J. 2007, Astron.\  Nachr., 328, 1002

\bibitem[]{Ci08} Cai, M. J., Shang, H., Lin, H.-H., \&
  Shu, F. H. 2008, \apj, 672, 489

\bibitem[]{Cv97} Calvet, N. 1997, in IAU Symp.\  182,
  Herbig-Haro Flows and the Birth of Stars, ed. B. Reipurth \&
  C. Bertout (Dordrecht: Kluwer), 417

\bibitem[]{CG98} Calvet, N., \& Gullbring, E. 1998, \apj, 509, 802

\bibitem[]{CH92} Calvet, N., \& Hartmann, L. 1992, \apj, 386, 239

\bibitem[]{Cm90} Camenzind, M. 1990, Rev.\  Mod.\  Astron., 3, 234

\bibitem[]{Css7} Casse, F., Meliani, Z., \& Sauty, C. 2007,
  \apss, 311, 57

\bibitem[]{Ch07} Chan, K. L. 2007, Astron.\  Nachr., 328, 1059

\bibitem[]{CI82} Chevalier, R. A., \& Imamura, J. N. 1982, \apj,
  261, 543

\bibitem[]{Cy66} Clayton, D. D. 1966, \aj, 71, 381

\bibitem[]{Co03} Cole, D. G. 2003, \ssr, 107, 295

\bibitem[]{Cr02} Cranmer, S. R. 2002, \ssr, 101, 229

\bibitem[]{Cr04} Cranmer, S. R. 2004, Am.\  J.\  Phys., 72, 1397

\bibitem[]{CvB05} Cranmer, S. R., \& van Ballegooijen, A. A. 2005,
  \apjs, 156, 265

\bibitem[]{CvB07} Cranmer, S. R., van Ballegooijen, A. A., \&
  Edgar, R. J. 2007, \apjs, 171, 520

\bibitem[]{Cu08} Cuntz, M., Gurdemir, L., Guinan, E. F., \& Kurucz,
  R. L. 2008, in IAU Symp.\  249, Exoplanets: Detection, Formation and
  Dynamics, ed. Y. Sun \& S. Ferraz-Mello (Cambridge: Cambridge
  Univ.\  Press), 203

\bibitem[]{DO73} Davidson, K., \& Ostriker, J. P. 1973, \apj, 179, 585

\bibitem[]{Dc81} DeCampli, W. M. 1981, \apj, 244, 124

\bibitem[]{De08} Delann\'{e}e, C., T\"{o}r\"{o}k, T., Aulanier, G.,
  \& Hochedez, J.-F. 2008, \solphys, 247, 123

\bibitem[]{dN87} de Jager, C., \& Nieuwenhuijzen, H. 1987, \aap,
  177, 217

\bibitem[]{DS83} Dennis, J. E., \& Schnabel, R. B. 1983,
  Numerical Methods for Unconstrained Optimization and Nonlinear
  Equations (Englewood Cliffs, NJ: Prentice-Hall)

\bibitem[]{Dm03} Dmitruk, P., \& Matthaeus, W. H. 2003, \apj,
  597, 1097

\bibitem[]{Dm02} Dmitruk, P., Matthaeus, W. H., Milano, L. J.,
  Oughton, S., Zank, G. P., \& Mullan, D. J. 2002, \apj, 575, 571

\bibitem[]{Dm01} Dmitruk, P., Milano, L. J., \& Matthaeus, W. H.
  2001, \apj, 548, 482

\bibitem[]{Dn07} Donati, J.-F., et al. 2007, \mnras, 380, 1297

\bibitem[]{Du05} Dupree, A. K., Brickhouse, N. S., Smith, G. H.,
  \& Strader, J. 2005, \apjl, 625, L131

\bibitem[]{Ed06} Edwards, S., Fischer, W., Hillenbrand, L., \&
  Kwan, J. 2006, \apj, 646, 319

\bibitem[]{Eg71} Eggleton, P. P. 1971, \mnras, 151, 351

\bibitem[]{Eg72} Eggleton, P. P. 1972, \mnras, 156, 361

\bibitem[]{Eg73} Eggleton, P. P. 1973, \mnras, 163, 279

\bibitem[]{Ee73} Eggleton, P. P., Faulkner, J., \& Flannery, B. P.
  1973, \aap, 23, 325

\bibitem[]{Ef04} Elfimov, A. G., Galv\~{a}o, R. M. O.,
  Jatenco-Pereira, V., \& Opher, R. 2004, \apj, 600, 292

\bibitem[]{EL77} Elsner, R. F., \& Lamb, F. K. 1977, \apj, 215, 897

\bibitem[]{Fg02} Falceta-Gon\c{c}alves, D., Vidotto, A. A., \&
  Jatenco-Pereira, V. 2006, \mnras, 368, 1145

\bibitem[]{FM99} Feigelson, E. D., \& Montmerle, T. 1999,
  \araa, 37, 363

\bibitem[]{Fe06} Ferreira, J., Dougados, C., \& Cabrit, S. 2006,
  \aap, 453, 785

\bibitem[]{FG00} Feynman, J., \& Gabriel, S. B. 2000, \jgr, 105, 10543

\bibitem[]{F65} Field, G. B. 1965, \apj, 142, 531

\bibitem[]{Fc3a} Flaccomio, E., Micela, G., \& Sciortino, S. 2003a,
  \aap, 397, 611

\bibitem[]{Fc3b} Flaccomio, E., Micela, G., \& Sciortino, S. 2003b,
  \aap, 402, 277

\bibitem[]{GS87} Gail, H. P., \& Sedlmayr, E. 1987, \aap, 171, 197

\bibitem[]{GA05} Garc\'{\i}a-Alvarez, D., Drake, J. J., Lin, L.,
  et al.  2005, \apj, 621, 1009

\bibitem[]{GL79a} Ghosh, P., \& Lamb, F. K. 1979a, \apj, 232, 259

\bibitem[]{GL79b} Ghosh, P., \& Lamb, F. K. 1979b, \apj, 234, 296

\bibitem[]{Gia07} Giardino, G., Favata, F., Pillitteri, I.,
  Flaccomio, E., Micela, G., \& Sciortino, S. 2007, \aap, 475, 891

\bibitem[]{GV07} G\'{o}mez de Castro, A. I., \& Verdugo, E. 2007,
  \apjl, 654, L91

\bibitem[]{Gv07} Grevesse, N., Asplund, M., \& Sauval, A. J. 2007,
  \ssr, 130, 105

\bibitem[]{Gd07} G\"{u}del, M. 2007, Living Rev.\  Solar Phys., 4,
  lrsp-2007-3

\bibitem[]{Gd98} G\"{u}del, M., Guinan, E. F., \& Skinner, S. L.
  1998, in 10th Cambridge Workshop on Cool Stars, Stellar Systems,
  and the Sun, ASP Conf.\  Ser.\  154, 1041

\bibitem[]{Gu96} Gullbring, E., Barwig, H., Chen, P. S., Gahm, G. F.,
  \& Bao, M. X. 1996, \aap, 307, 791

\bibitem[]{Hm82} Hammer, R. 1982, \apj, 259, 767

\bibitem[]{Hn94} Han, Z., Podsiadlowski, P., \& Eggleton, P. P.
  1994, \mnras, 270, 121

\bibitem[]{HEG95} Hartigan, P., Edwards, S., \& Ghandour, L. 1995,
  \apj, 452, 736

\bibitem[]{HEG04} Hartigan, P., Edwards, S., \& Pierson, R. 2004,
  \apj, 609, 261

\bibitem[]{H00} Hartmann, L. 2000, Accretion Processes in Star
  Formation (Cambridge: Cambridge Univ.\  Press)

\bibitem[]{HC98} Hartmann, L., Calvet, N., Gullbring, E., \&
  D'Alessio, P. 1998, \apj, 495, 385

\bibitem[]{HC97} Hartmann, L., Cassen, P., \& Kenyon, S. J. 1997,
  \apj, 475, 770

\bibitem[]{HHC94} Hartmann, L., Hewett, R., \& Calvet, N. 1994,
  \apj, 426, 669

\bibitem[]{HM80} Hartmann, L., \& MacGregor, K. B. 1980,
  \apj, 242, 260

\bibitem[]{HO80} Heinemann, M., \& Olbert, S. 1980, \jgr,
  85, 1311

\bibitem[]{Hf05} H\"{o}fner, S. 2005, in 13th Cambridge Workshop on
  Cool Stars, Stellar Systems, and the Sun, ed. F. Favata, G. Hussain,
  \& B. Battrick (Noordwijk, The Netherlands: ESA), ESA SP-560, 335 

\bibitem[]{Ho74} Hollweg, J. V. 1974, \jgr, 79, 3845

\bibitem[]{Ho76} Hollweg, J. V. 1976, \jgr, 81, 1649

\bibitem[]{Ho86} Hollweg, J. V. 1986, \jgr, 91, 4111

\bibitem[]{HA70} Holzer, T. E., \& Axford, W. I. 1970, \araa, 8, 31

\bibitem[]{HFL83} Holzer, T. E., {Fl\aa}, T., \& Leer, E. 1983,
  \apj, 275, 808

\bibitem[]{Hz07} Holzwarth, V. 2007,
  Mem.\  Soc.\  Astron.\  Italiana, 78, 271

\bibitem[]{HJ07} Holzwarth, V., \& Jardine, M. 2007, \aap, 463, 11

\bibitem[]{Hs95} Hossain, M., Gray, P. C., Pontius, D. H., Jr.,
  Matthaeus, W. H., \& Oughton, S. 1995, Phys.\  Fluids,
  7, 2886

\bibitem[]{IS75} Illarionov, A. F., \& Sunyaev, R. A. 1975, \aap,
  39, 185

\bibitem[]{J77} Jacques, S. A. 1977, \apj, 215, 942

\bibitem[]{Jr08} Jardine, M. M., Gregory, S. G., \& Donati, J.-F.
  2008, \mnras, 386, 688

\bibitem[]{Jf06} Jeffries, R. D., Evans, P. A., Pye, J. P., \&
  Briggs, K. R. 2006, \mnras, 367, 781

\bibitem[]{JB95} Johns, C. M., \& Basri, G. 1995, \apj, 449, 341

\bibitem[]{Ju03} Judge, P. G., Solomon, S. C., \& Ayres, T. R.
  2003, \apj, 593, 534

\bibitem[]{Ka07} K\"{a}pyl\"{a}, P. J., Korpi, M. J., Stix, M., \&
  Tuominen, I. 2007, in IAU Symp.\  239, Convection in Astrophysics,
  ed. F. Kupka, I. Roxburgh \& K. Chan (Cambridge: Cambridge
  Univ.\  Press), 437

\bibitem[]{Ki02} Killie, M. A. 2002, M.S.\  Thesis, University of
  Oslo, Norway

\bibitem[]{Kl06} Klimchuk, J. A. 2006, \solphys, 234, 41

\bibitem[]{K06} Kohl, J. L., Noci, G., Cranmer, S. R., \&
  Raymond, J. C. 2006, \aapr, 13, 31

\bibitem[]{Ko91} K\"{o}nigl, A. 1991, \apjl, 370, L39

\bibitem[]{KP00} K\"{o}nigl, A., \& Pudritz, R. E. 2000, in
  Protostars and Planets IV, ed. V. Mannings, A. Boss, S. Russell
  (Arizona: Univ.\  Arizona Press), 759

\bibitem[]{Ku08} Kulkarni, A. K., \& Romanova, M. M. 2008,
  \mnras, 386, 673

\bibitem[]{KIS} Kuperus, M., Ionson, J. A., \& Spicer, D. S.
  1981, \araa, 19, 7

\bibitem[]{K92} Kurucz, R. L. 1992, in IAU Symp.\  149,
  The Stellar Populations of Galaxies, ed. B. Barbuy \& A. Renzini
  (Dordrecht: Kluwer), 225

\bibitem[]{Ld85} Lada, C. J. 1985, \araa, 23, 267

\bibitem[]{Lz99} Lamzin, S. A. 1999, Astron.\  Lett., 25, 430

\bibitem[]{Lt98} Lee, T., Shu, F. H., Shang, H., Glassgold, A. E., \&
  Rehm, K. E. 1998, \apj, 506, 898

\bibitem[]{Le98} Leer, E., Hansteen, V. H., \& Holzer, T. E.
  1998, in Cyclical Variability in Stellar Winds, ed.
  L. Kaper \& A. W. Fullerton (Berlin: Springer-Verlag), 263

\bibitem[]{Lr07} Long, M., Romanova, M. M., \& Lovelace, R. V. E.
  2007, \mnras, 374, 436

\bibitem[]{L71} Lucy, L. B. 1971, \apj, 163, 95

\bibitem[]{L76} Lucy, L. B. 1976, \apj, 205, 482

\bibitem[]{LP74} Lynden-Bell, D., \& Pringle, J. E. 1974,
  \mnras, 168, 603

\bibitem[]{MP05} Matt, S., \& Pudritz, R. E. 2005, \apjl, 632, L135

\bibitem[]{MP07} Matt, S., \& Pudritz, R. E. 2007,
  in IAU Symp.\  243, Star-Disk Interactions in Young Stars, ed. J.
  Bouvier, I. Appenzeller (Cambridge: Cambridge U.\  Press), 299

\bibitem[]{MP08} Matt, S., \& Pudritz, R. E. 2008, \apj, 681, 391

\bibitem[]{MZO} Matthaeus, W. H., Zank, G. P., Oughton, S.,
  Mullan, D. J., \& Dmitruk, P. 1999, \apjl, 523, L93

\bibitem[]{Mz98} Mazzotta, P., Mazzitelli, G., Colafrancesco, S.,
  \& Vittorio, N. 1998, \aaps, 133, 403

\bibitem[]{McOs} McKee, C. F., \& Ostriker, E. C. 2007, \araa,
  45, 565

\bibitem[]{Mn05} Mignone, A. 2005, \apj, 626, 373

\bibitem[]{M78} Mihalas, D. 1978, Stellar Atmospheres, 2nd ed.
  (San Francisco: W. H. Freeman)

\bibitem[]{Mb04} Montalb\'{a}n, J., D'Antona, F., Kupka, F., \&
  Heiter, U. 2004, \aap, 416, 1081

\bibitem[]{MR60} Moreton, G. E., \& Ramsey, H. E. 1960, \pasp,
  72, 357

\bibitem[]{MC93} Mullan, D. J., \& Cheng, Q. Q. 1993, \apj, 412, 312

\bibitem[]{Mz02} Musielak, Z. E., \& Ulmschneider, P. 2002,
  \aap, 386, 606

\bibitem[]{Mu00} Muzerolle, J., Calvet, N., Brice\~{n}o, C.,
  Hartmann, L., \& Hillenbrand, L. 2000, \apjl, 535, L47

\bibitem[]{Mu01} Muzerolle, J., Calvet, N., \& Hartmann, L. 2001,
  \apj, 550, 944

\bibitem[]{NU90} Narain, U., \& Ulmschneider, P. 1990, \ssr,
  54, 377

\bibitem[]{NU96} Narain, U., \& Ulmschneider, P. 1996, \ssr,
  75, 453

\bibitem[]{Ny84} Noyes, R. W., Hartmann, L. W., Baliunas, S. L.,
  Duncan, D. K., \& Vaughan, A. H. 1984, \apj, 279, 763

\bibitem[]{OD06} Oughton, S., Dmitruk, P., \& Matthaeus, W. H.
  2006, Phys.\  Plasmas, 13, 042306

\bibitem[]{Ow04} Owocki, S. P. 2004, in Evolution of Massive Stars,
  Mass Loss, and Winds, ed. M. Heydari-Malayeri, P. Stee, \&
  J.-P. Zahn, EAS Pub.\  Ser. 13, 163

\bibitem[]{OM08} Owocki, S. P., \& van Marle, A. J. 2008, in
  IAU Symp.\  250, Massive Stars as Cosmic Engines, ed. F. Bresolin,
  P. Crowther, J. Puls (Cambridge: Cambridge U.\  Press), 71

\bibitem[]{PC96} Paatz, G., \& Camenzind, M. 1996, \aap, 308, 77

\bibitem[]{P53} Parker, E. N. 1953, \apj, 117, 431

\bibitem[]{P58} Parker, E. N. 1958, \apj, 128, 664

\bibitem[]{P63} Parker, E. N. 1963, Interplanetary Dynamical
  Processes (New York: Interscience)

\bibitem[]{Po95} Pols, O. R., Tout, C. A., Eggleton, P. P., \&
  Han, Z. 1995, \mnras, 274, 964

\bibitem[]{PF05} Preibisch, T., \& Feigelson, E. D. 2005, \apjs,
  160, 390

\bibitem[]{Rx76} Raymond, J. C., Cox, D. P., \& Smith, B. W. 1976,
  \apj, 204, 290

\bibitem[]{Rm75} Reimers, D. 1975,
  Mem.\  Soc.\  R.\  Sci.\  Li\`{e}ge, 8, 369

\bibitem[]{Rm77} Reimers, D. 1977, \aap, 61, 217

\bibitem[]{Rz77} Renzini, A., Cacciari, C., Ulmschneider, P., \&
  Schmitz, F. 1977, \aap, 61, 39

\bibitem[]{Ri05} Ribas, I., Guinan, E. F., G\"{u}del, M., \& Audard, M.
  2005, \apj, 622, 680

\bibitem[]{Rb04} Robinson, F. J., Demarque, P., Li, L. H., Sofia, S.,
  Kim, Y.-C., Chan, K. L., \& Guenther, D. B. 2004, \mnras, 347, 1208

\bibitem[]{Rm04} Romaniello, M., Robberto, M., \& Panagia, N. 2004,
  \apj, 608, 220

\bibitem[]{Rv08} Romanova, M. M., Kulkami, A. K., \& Lovelace,
  R. V. E. 2008, \apjl, 673, L171

\bibitem[]{Rs91} Rosner, R., An, C.-H., Musielak, Z. E., Moore, R. L.,
  \& Suess, S. T. 1991, \apjl, 372, L91

\bibitem[]{Rs95} Rosner, R., Musielak, Z. E., Cattaneo, F., Moore,
  R. L., \& Suess, S. T. 1995, \apjl, 442, L25

\bibitem[]{RTV} Rosner, R., Tucker, W. H., \& Vaiana, G. S. 1978,
  \apj, 220, 643

\bibitem[]{Sa01} Saar, S. H. 2001, in Cool Stars, Stellar Systems,
  and the Sun, 11th Cambridge Workshop, ed. R. Garcia Lopez, R.
  Rebolo, \& M. Zapaterio Osorio, ASP Conf.\  Ser.\  223, 292

\bibitem[]{Sf98} Safier, P. N., 1998, \apj, 494, 336

\bibitem[]{SK88} Scheurwater, R., \& Kuijpers, J. 1988, \aap,
  190, 178

\bibitem[]{SC05} Schr\"{o}der, K.-P., \& Cuntz, M. 2005, \apjl,
  630, L73

\bibitem[]{Sw06} Schwadron, N. A., McComas, D. J., \& DeForest,
  C. 2006, \apj, 642, 1173

\bibitem[]{Sh07} Shu, F. H., Galli, D., Lizano, S., \& Cai, M. J. 2007,
  in IAU Symp.\  243, Star-Disk Interactions in Young Stars, ed. J.
  Bouvier, I. Appenzeller (Cambridge: Cambridge U.\  Press), 249

\bibitem[]{Sk72} Skumanich, A. 1972, \apj, 171, 565

\bibitem[]{Sm03} Stempels, H. C. 2003, Ph.D.\  Dissertation,
  Uppsala Universitet, Sweden

\bibitem[]{Sk89} Strom, K. M., Strom, S. E., Edwards, S., Cabrit, S.,
  \& Skrutskie, M. F. 1989, \aj, 97, 1451

\bibitem[]{St04} Struck, C., Smith, D. C., Willson, L. A., Turner, G.,
  \& Bowen, G. H. 2004, \mnras, 353, 559

\bibitem[]{Su06} Suzuki, T. K. 2006, \apjl, 640, L75

\bibitem[]{Su07} Suzuki, T. K. 2007, \apj, 659, 1592

\bibitem[]{Te07} Telleschi, A., G\"{u}del, M., Briggs, K. R.,
  Audard, M., \& Palla, F. 2007, \aap, 468, 425

\bibitem[]{Th98} Thompson, B. J., Plunkett, S. P., Gurman, J. B.,
  Newmark, J. S., St. Cyr, O. C., \& Michels, D. J. 1998, \grl,
  25, 2465

\bibitem[]{TM95} Tu, C.-Y., \& Marsch, E. 1995, \ssr, 73, 1

\bibitem[]{U68} Uchida, Y. 1968, \solphys, 4, 30

\bibitem[]{US84} Uchida, Y., \& Shibata, K. 1984, \pasj, 36, 105

\bibitem[]{UM86} Ulmschneider, P., \& Muchmore, D. 1986, in
  Small Scale Magnetic Flux Concentrations in the Solar
  Photosphere, ed.  W. Deinzer, M. Kn\"{o}lker, \& H. H. Voigt
  (G\"{o}ttingen: Vandenhoeck \& Ruprecht), 191

\bibitem[]{Us06} Ustyugova, G. V., Koldoba, A. V., Romanova, M. M.,
  \& Lovelace, R. V. E. 2006, \apj, 646, 304

\bibitem[]{vB94} van Ballegooijen, A. A. 1994, \ssr, 68, 299

\bibitem[]{Vc02} Vasconcelos, M. J., Jatenco-Pereira, V., \&
  Opher, R. 2002, \apj, 574, 847

\bibitem[]{Ve01} Velli, M. 2001, \apss, 277, 157

\bibitem[]{Wy00} Wang, Y.-M. 2000, \apjl, 543, L89

\bibitem[]{WS91} Wang, Y.-M., \& Sheeley, N. R., Jr. 1991,
  \apjl, 372, L45

\bibitem[]{Wh97} Whang, Y. C. 1997, \apj, 485, 389

\bibitem[]{WD07} Wills-Davey, M. J., DeForest, C. E., \& Stenflo,
  J. O. 2007, \apj, 664, 556

\bibitem[]{Wi00} Willson, L. A. 2000, \araa, 38, 573

\bibitem[]{Wt88} Withbroe, G. L. 1988, \apj, 325, 442

\bibitem[]{Wo02} Wood, B. E., M\"{u}ller, H.-R., Zank, G. P., \&
  Linsky, J. L. 2002, \apj, 574, 412

\bibitem[]{Wo05} Wood, B. E., M\"{u}ller, H.-R., Zank, G. P.,
  Linsky, J. L., \& Redfield, S. 2005, \apjl, 628, L143

\bibitem[]{ZM90} Zhou, Y., \& Matthaeus, W. H. 1990, \jgr,
  95, 10291

\end{thebibliography}
\end{document}